\documentclass[]{aa}

\usepackage{graphicx}
\usepackage{txfonts}
\usepackage[normalem]{ulem}
\usepackage[dvipsnames]{xcolor}
\begin{document}

   \title{Uncovering a new group of T Tauri stars in the Taurus-Auriga molecular complex from \textit{Gaia} and GALEX data\footnote{Table 2 is only available in electronic form at the CDS via anonymous ftp to cdsarc.cds.unistra.fr (130.79.128.5) or via \url{https://cdsarc.cds.unistra.fr/cgi-bin/qcat?J/A+A/}}}

   \subtitle{}

   \author{Ana In\'es G\'omez de Castro
          \inst{1,2 },  Ra\'ul de la Fuente Marcos\inst{1,2 }, Ada Canet \inst{1,2 }, Leire Beitia-Antero, \inst{1,3}, Javier Y\'a\~nez-Gestoso \inst{1, 3}, 
          \and 
          Juan Carlos Vallejo\inst{1,2 }
          }

   \institute{AEGORA Research Group - Joint Center for Ultraviolet Astronomy, Universidad Complutense de Madrid, Plaza de Ciencias 3, 28040 Madrid, Spain
              \email{}
              \and
                Departamento de Fisica de la Tierra y Astrofisica, Fac. de CC. Matematicas, Plaza de Ciencias 3, 28040 Madrid, Spain
                \and 
                Departamento de Estad\'{\i}stica e Investigaci\'on Operativa, Fac. de CC. Matem\'aticas, Plaza de Ciencias 3, 28040 Madrid, Spain
                \\
             \email{aig@ucm.es}
             \thanks{}
             }

   \date{Received ; accepted }

 
  \abstract
   {The determination of the complete census of young stars in any star forming region is a challenge even for the nearest and best observed molecular clouds such as Taurus-Auriga (TAMC). Deep surveys at infrared (IR) and X-ray wavelengths and astrometric surveys using \textit{Gaia}  DR2 and DR3  have been carried out to detect the sparse population and constrain the low mass end of the initial mass function. These compilations have resulted in lists of  more than 500  sources including reliable members of the association and candidates.The astrometric information provided by  the \textit{Gaia} mission has proven fundamental to evaluate these candidates.}
   {In this work, we examine the list of 63 candidates to T Tauri star (TTS) in the TAMC identified by their ultraviolet (UV) and infrared colours (IR) measured from data obtained by the Galaxy Evolution Explorer all sky survey (GALEX-AIS) and the Two Microns All Sky Survey (2MASS), respectively. These sources have not been included in previous studies and the objective of this work is twofold: evaluate whether they are pre-main sequence (PMS) stars and  evaluate the goodness of the UV-IR colour-colour diagram to detect PMS stars in wide-fields.}
   {The kinematic properties and the parallax of these sources have been retrieved from the \textit{Gaia} DR3 catalogue and used to evaluate their membership probability. Several classification algorithms have been tested to search for the kinematical groups but the final classification has been made with  $k$-means$++$ algorithms. Membership probability has been evaluated by applying Logistic Regression. In addition, spectroscopic information available in the archive of the Large Sky Area Multi Object Fiber Spectroscopic Telescope (LAMOST) has been used to ascertain their PMS nature when available. }
   {About 20\% of the candidates share the kinematics of  the TAMC members. Among them, HD 281691 is a G8-type field star located in front of the cloud and HO Aur is likely a halo star given the very low metallicity provided by  \textit{Gaia}. The rest are three known PMS stars (HD 30171, V600 Aur and J04590305$+$3003004), two previously unknown accreting M-type stars (J04510713$+$1708468 and J05240794$+$2542438) and, five additional sources, which are very likely PMS stars. Most of these new sources are concentrated at low galactic latitudes over the Auriga-Perseus region.   }
   {}


\titlerunning{New TTSs in the Taurus-Auriga MC}
\authorrunning{Gomez de Castro et al.}

   \maketitle
%

\section{Introduction}

The accurate determination of the low-mass end of the initial mass function (IMF)  is challenging but fundamental to understand the efficiency of star formation and to constrain the significance of the low-mass leftovers in the evolution of galaxies over cosmic scales. There have been systematic attempts to make a complete accounting of  this "hidden" population in the solar neighbourhood where the sensitivity for this investigation  is optimal (see {\it i.e} \citealt{2018AJ....156..271L} and references therein).  The nearest stellar nurseries are located within 200 pc, in a rim of molecular clouds that includes prominent star forming regions such as Taurus-Auriga, Lupus, Chamaleon,  Ophiuchus or Sco-Cen. As a result, the Solar System is in a privileged location within the Galaxy to investigate the formation of loose stellar associations and determine the low mass end of their IMF.

  \begin{figure*}
   \centering
   \includegraphics[width=16cm]{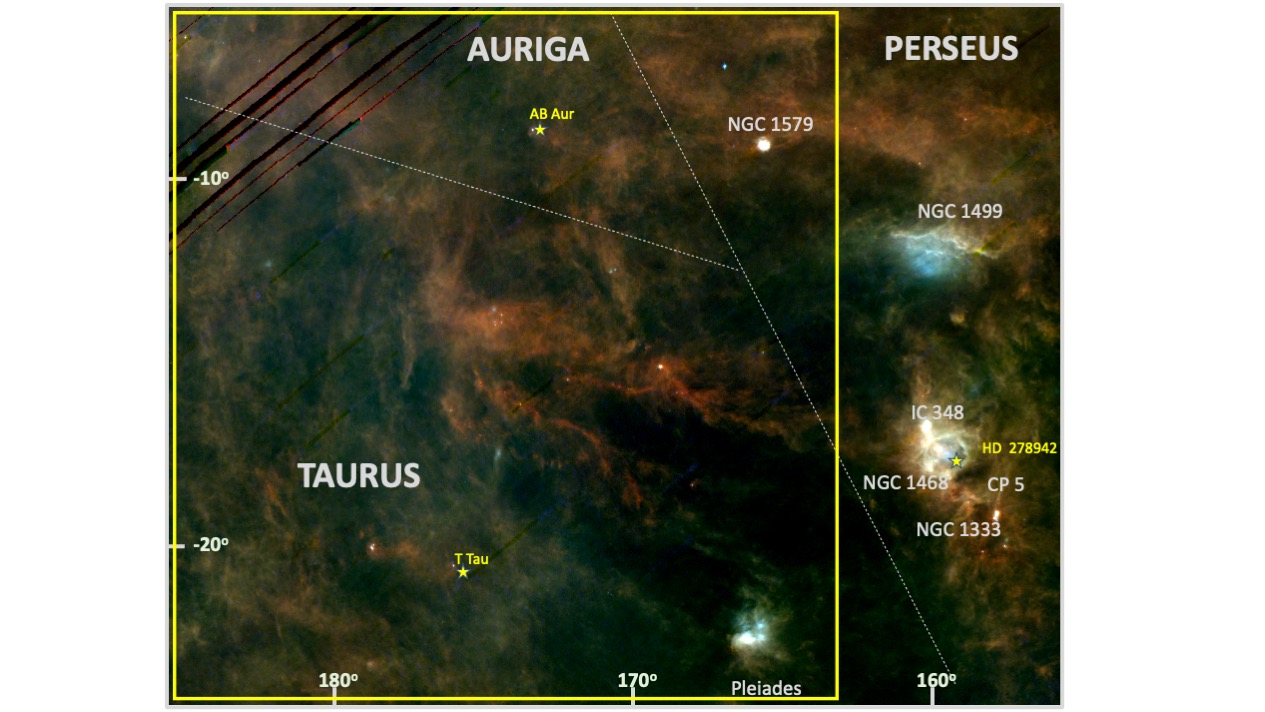}
      \caption{Map of the dust distribution in the Taurus, Auriga, Perseus star forming complexes obtained by the Far Infrared Surveyor
      on board the AKARI satellite. The location of the main young associations is indicated as well as that of T Tau and AB Aur, two
      prominent pre-main sequence stars in the region. The yellow frame marks the area surveyed by GALEX and searched
      for young stars by  GdC2015. 
              }
         \label{akari}
   \end{figure*}

The Taurus-Auriga molecular complex (TAMC)  is the best studied region in the area.  The TAMC  is part of the  prominent ridge of molecular gas observed towards the anticenter of the Galaxy; large-scale maps of the region date back to the CO survey carried with the Columbia millimeter-wave telescope by \citet{1987ApJS...63..645U}.  The TAMC roughly extends  over $20^{\rm o} \times 20^{\rm o}$ in the sky (see Fig. \ref{akari}) and it is constituted by several filaments and cores 
\citep{2008ApJ...680..428G, 2021A&A...645A.145G, 2020A&A...638A..85R, 2019ApJ...882...49L}. There are more than 500 sources sparsely distributed over the complex that could be pre-main sequence (PMS) stars (see \citealt{2022arXiv221109785L} for a recent accounting). The very young population (age $\leq 1 $Myr)  is well identified by its significant  near-infrared excess, its proximity to the parent molecular gas and the presence of jets of molecular outflows; however, the census of the most evolved sources, in particular, the  so-called weak line T Tauri stars (WTTSs), is still to be completed, especially at the very low mass end towards the brown dwarfs frontier. 

The release of the \textit{Gaia} DR2 catalogue \citep{2018A&A...616A...1G} enabled for the first time an unbiased search for WTTSs in the region based on the  kinematical properties of the stars. The \textit{Gaia} survey is magnitude-limited to $G\simeq 21$ and thus, it is sensitive to the low mass stellar population (late M types) in the TAMC which is located at  $\sim 140$ pc. \citet{2019A&A...630A.137G}
generated a compilation of  all the possible T Tauri stars (TTSs) in the region including both, confirmed spectroscopic sources and possible candidates, which were identified by the surveys carried out with the Spitzer  infrared telescope \citep{2010ApJS..186..111L}, the Sloan Digital Sky Survey and the 2-Microns All Sky Survey \citep{2017AJ....153...46L}.  Using a Hierarchical Mode Association Clustering (HMAC) algorithm, they classified these sources into 21 distinct kinematical groups.  A recent re-evaluation based on \textit{Gaia} DR3 
\citep{2022arXiv220800211G}
data has resulted in a grand total of 532 {\it adopted members} distributed in 13 kinematical groups \citep{2022arXiv221109785L}.  
This kinematical search does not provide a complete accounting of the young stars in the region since young stars with peculiar proper motions
such as  stars ejected in close encounters or stars in multiple systems, will be rejected as outliers by the classification algorithms. Also, 
no information on their youth and evolutionary state can be obtained in this manner. An additional problem is the scarce and uncertain information provided by \textit{Gaia} 
DR3 on the radial velocity of these stars which makes difficult  an accurate  determination of the three components of the LSR velocity $(U,V,W)$ for a large fraction of the sources. 

The most common and successful ways to search for WTTSs in molecular clouds are based on their strong magnetic activity which results in an enhancement of the atmospheric flux: coronal emission at X-ray, the radiation from the transition region at ultraviolet (UV) wavelengths and the chromospheric radiation measurable at UV but also at optical wavelengths. In fact, the first catalogues of TTSs were built from stars displaying enhanced chromospheric emission at the H$\alpha$ line 
(see {\it e.g.} \citealt{1972ApJ...174..401H}).  Many WTTSs candidates were identified in the TAMC from the ROSAT all-sky X-ray survey  \citep{1995A&A...297..391N} and subsequent observations (especially of the Li I absorption) proved that many were indeed, WTTSs. The release of the {\it "all sky UV survey"} carried out by the Galaxy Evolution Explorer (GALEX) enabled to search for TTSs by their UV excess; a list of 63 new TTSs candidates was elaborated by comparing the UV color and the IR color excesses  in the TAMC (\citealt{2015ApJS..216...26G}, hereafter GdC2015). \textit{Gaia} was launched in December 2013 and its subsequent data releases came much later than the publication of this list thus, neither parallax nor kinematical information
was available at the time to narrow down this large list of candidates. It is the purpose of this research  to carry out such analysis, to identify the {\it bona fide} TTSs in the sample. For this purpose, we use the \textit{Gaia} DR3 parallax and kinematical information as well as spectra from the spectroscopic survey carried out by the Large Sky Area Multi-Object Fiber Spectroscopic Telescope (LAMOST)  when available, to determine the spectral type and the Li I equivalent width. 

This article is structured as follows. In section 2, the characteristics of the GdC2015 sources are described.  Sources with \textit{Gaia} DR3 parallaxes incompatible with TAMC membership are disregarded reducing the working sample of viable candidates to 13 sources. In section 3, the membership probability  of these candidates is evaluated using \textit{Gaia} DR3 data. In the process, a list of {\it trusted TTSs} in the TAMC for kinematical studies is produced; this compilation is significantly shorter than that produced by \citet{2022arXiv221109785L}. In section 4, LAMOST spectra are analyzed; two new accreting late M-type stars are identified in the GdC2015 sample.  Finally,  a short discussion (Sect. 5) on the radial velocity distribution of the TTSs and the coupling with the molecular gas is included.  A short summary of the work is at the end, in Section 6.  Technicalities concerning the error analysis of the samples are compiled in the Appendix to facilitate a comprehensive reading of the work.

\section{UV candidates to T Tauri stars in the Taurus-Auriga region}

  \begin{figure}[h!]
   \centering
   \includegraphics[width=9 cm]{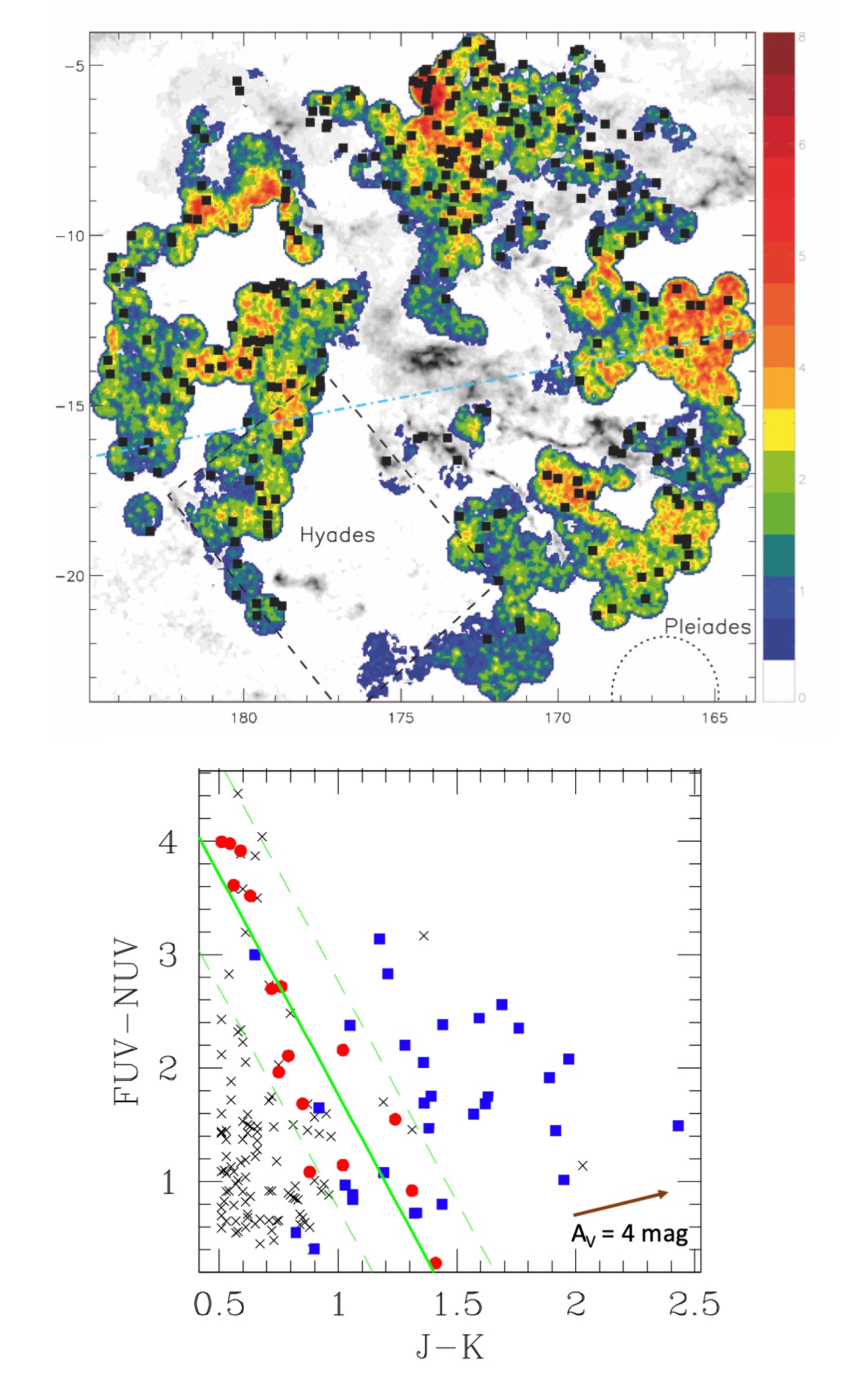}
      \caption{Candidates to TTSs in the TAMC from GdC2015. {\it Top panel}, the location of the TTSs and candidates on the sky is overlaid on the
      density of GALEX NUV sources, in galactic coordinates; stellar  densities are color coded. The density of molecular gas  is outlined
      from the 2MASS extinction map by \citealt{2010A&A...512A..67L}).  {\it Bottom panel}, color-color diagram used for the selection of the TTSs candidates in GdC2015. Candidates are marked with black crosses, and known CTTSs and WTTSs from the qualification sample are represented by blue squares and red circles, respectively. The regression line marking the location of the WTTSs in the diagram is plotted (solid green line) as well as the uncertainty band from the fit (dashed green lines); most of the TTSs
      candidates to be analyzed in this work are within this strip.  }
         \label{gdc2015}
   \end{figure}

The GALEX All Sky Survey (AIS)  \citep{2005ApJ...619L...1M}  provided for the first time an unbiased view of the TAMC at UV wavelengths. Though the 
area was not mapped completely due to GALEX sensitivity constraints, 380 square degrees in the sky (equivalent to 197 GALEX  fields)  
were imaged  and as many as $\sim$163,000  UV sources  were detected  in the near ultraviolet (NUV) band (see  GdC2015). Through a 
comparison of UV and IR colours, 63 new candidates to TTSs were identified; their location is shown in the top panel of Fig. \ref{gdc2015}
 and their UV-IR colours in the bottom panel. 
 
These 63 sources were selected using a qualification sample of TTSs that was generated from two sets of sources:
[1] the 31 known TTSs detected in the GALEX AIS survey of the TAMC and [2] the 21 TTSs observed with the  International Ultraviolet Explorer (IUE) in the low dispersion mode  with high signal-to-noise ratio spectra (see Table~1 in GdC2015); synthetic GALEX photometry was calculated from the IUE spectra for these sources using Morrisey et al. (2007) conversion. 2MASS photometry was available for all the sources in the qualification sample. 
All these stars are marked with blue squares (if CTTSs) and red circles (if WTTSs) in the bottom panel of Fig. \ref{gdc2015}. 
The 63 candidates to be evaluated in this work appear as crosses in the plot.

 Note that the WTTSs define a  clear regression line with the hardest UV colours being observed in the sources displaying the smallest
J-K color. CTTSs however, are scattered in the diagram.  Most of the candidates studied in this work were identified in the WTTSs strip (see  Fig. 2, bottom panel) 
close to the location of reddened massive cool stars and many are located close to Galactic plane (see Fig. 2, top panel). Thus, further observables such as distance  or  kinematics are needed to confirm the PMS nature of these sources.  

\textit{Gaia} DR3 counterparts have been searched and found for all candidates using a 2 arcsecond search radius employing the 
cross match services provided by the Centre de Donn\'ees Stellaires (CDS). 
 
   \begin{figure*}[h!]
   \centering
   \includegraphics[width=13 cm]{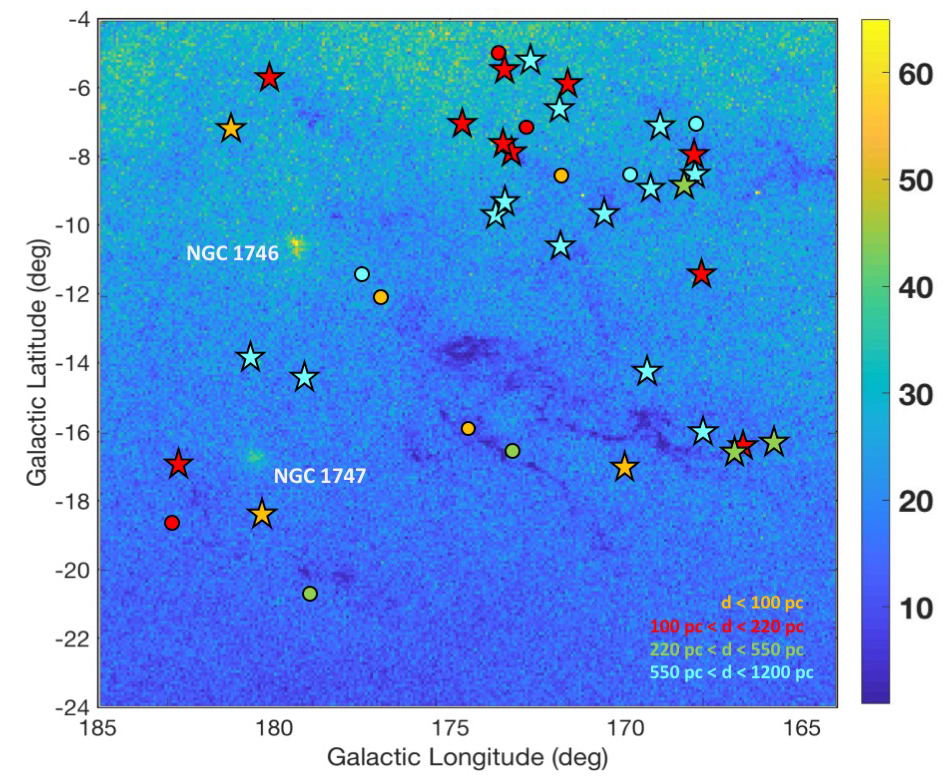}
      \caption{Density of \textit{Gaia} sources in the TAMC. The densities are color coded  in stars per 6 arcmin$^2$ (see lateral bar).  The
      shadow produced by the TAMC filaments over the stellar background is readily identified. The two over dense regions correspond
      to NGC 1747 (an open cluster at 550 pc) and NGC 1746, an asterism or apparent concentration of stars 
      \citep{2020A&A...633A..99C}, are easily identifiable.  The location of the GdC2015 candidates with \textit{Gaia} counterparts within 1200 pc is marked by circles and asterisks, color coded by distance. Asterisks represent  sources with high quality \textit{Gaia} measurements (RUWE $\leq 1.4$) and circles sources with RUWE larger than this threshold.
              }
         \label{gaia}
   \end{figure*}

In Fig. \ref{gaia}, we display the projected location on the plane of the sky of all sources within 1,200 pc. Only 13 among them have parallaxes compatible with TAMC  membership; {\it i.e.} 5 mas $-$ 9 mas, according to \citet{2020A&A...638A..85R}.  Seven among them are concentrated in the north of the region, close to the Galactic plane,  but are separated in two distance groups at $\sim 110$ pc and $\sim 160$ pc. 

\section{Kinematical properties of the candidates}

In addition to distance, stellar kinematics needs to be analyzed. This has been done in two steps. 
Firstly, the kinematics of known and trusted TTSs in the TAMC with high quality \textit{Gaia} measurements is studied.\footnote{\url{https://gea.esac.esa.int/archive/documentation/GDR2/Gaia_archive/chap_datamodel/sec_dm_main_tables/ssec_dm_ruwe.html}} Kinematical groups are identified within this sample using the
$k$-means ++  algorithm and then, the probability of the 13 candidates being kinematical members is  evaluated. Only trusted PMS sources with high 
quality astrometric data will be considered for this qualification sample. The \textit{Gaia} Data Processing and Analysis Consortium states that a value of the re-normalised
unit weight error (RUWE) $\le$1.4 is indicative of a good astrometric solution.\footnote{\url{https://gea.>
el/sec_dm_main_tables/ssec_dm_ruwe.html [gea.esac.esa.int]}} thus, only \textit{Gaia} observations with RUWE $\le$1.4 have been considered
for the qualification of the kinematics of the TTSs in the TAMC. As an aside, note that according to \citet{2021ApJ...907L..33S}  RUWE values
even slightly above 1.0 may be signaling unresolved binaries in \textit{Gaia} data.

\subsection{Kinematics of known TTSs in the TAMC}

The initial sample of trustable TTSs in the TAMC has been taken from \citet{2017A&A...599A..14J}; more recent catalogues are available, {\it e.g.}
\citet{2019A&A...630A.137G} and \citet{2022arXiv221109785L}, but they include together with {\it bona fide} TTSs, candidates to TTSs from various surveys.
\citet{2017A&A...599A..14J} compilation includes 338 sources and it is mainly based in the mid-infrared survey of the TAMC carried out with the {\it Spitzer}
telescope to perform a census of the disc population in the region that was completed with H$\alpha$ data for the few TTSs that were not detected
\citep{2010ApJS..186..111L}; only 274 sources of the catalogue have GAIA DR3 counterpart and only 170 have measurements with  RUWE $ \leq 1.4$.
 
Thus,  the initial qualification sample consisted of these 170 TTS with high quality astrometric measurements; however, after cross-checking \textit{Gaia} parallaxes three sources were found 
no to belong to the TAMC and have not been considered (see Table \ref{table:1}).  The final qualification sample consists of 92 classical TTSs (CTTSs) 
and 75 weak-line TTSs (WTTSs)  (see Table \ref{table:2}). 

\begin{table*}
\caption{Sources misclassified as TTSs in the TAMC}           
\label{table:1}    
\centering                         
\begin{tabular}{c c c c}        
\hline\hline                 
Name & RUWE& Parallax(mas)& Distance \\    
\hline         
2MASS J04375670$+$2546229  &         1.056 & $0.9734 \pm 0.0877$ & $\sim 1030$ pc \\
2MASS J04163048$+$3037053 &          1.003& $3.0552 \pm 0.094$ & $317-338$ pc \\
2MASS J04080782$+$2807280 &          1.201& $4.3569 \pm 0.0543$ & $227-232$ pc \\
\hline                                   
\end{tabular}
\end{table*}

\begin{table*}
\caption{Qualification Sample for the Kinematics of the TTSs in the TAMC$^{*}$}           
\label{table:2}    
\begin{flushleft}  
\begin{tiny}                   
\begin{tabular}{l l l l l l l l l l}      
\hline\hline                 
Name & l$_{\rm gal}$ & b$_{\rm gal}$ & Parallax & Parallax Err. & PM (RA) & err. PM(RA) &  PM (DEC) & err. PM(DEC) & RUWE \\
  & (deg) & (deg) & mas & mas & mas yr$^{-1}$ & mas yr$^{-1}$ &mas yr$^{-1}$ &mas yr$^{-1}$ & \\
\hline         
$2MASS J04161210+2756385$ &168.728127 &-16.2045517&	7.426 &0.0989 & 8.735 &	0.139 &-24.99	&0.089 &	0.88\\
$2MASS J04202555+2700355$ &170.105104 &-16.1639581&	6.0362&0.1102	&11.175&	0.132&-17.696 &0.09& 0.921\\
\hline                                   
\end{tabular}
\end{tiny}
\begin{tabular}{ll}
$^*$ & Full table is available on-line \\
\end{tabular}
\end{flushleft}  
\end{table*}

\begin{figure}
   \centering
   \includegraphics[width=9cm]{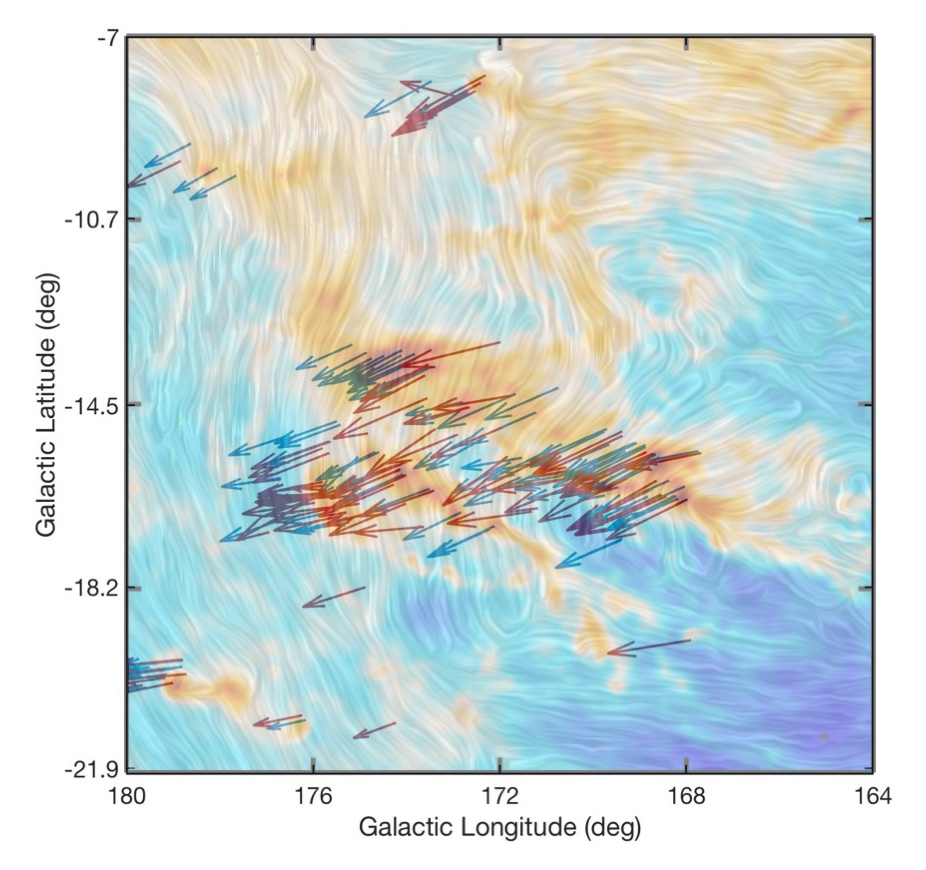}
      \caption{Proper motions (velocity projected in the plane of the sky) of the known TTSs in the TAMC with high quality astrometric measurements ($RUWE < 1.4$). CTTSs and WTTSs are represented with blue and red arrows, respectively.  Note that the motion of the stars is not parallel to the 
      filaments and it is approximately perpendicular to the direction of the magnetic field.     }
         \label{fig4}
   \end{figure}

\textit{Gaia} proper motions in the ICRS system have been converted into the Galactic system using the procedure described
in the \textit{Gaia} EDR3 documentation.\footnote{\url{https://gea.esac.esa.int/archive/documentation/GEDR3/Data_processing/chap_cu3ast/sec_cu3ast_intro/ssec_cu3ast_intro_tansforms.html}}
In Fig  \ref{fig4}, these proper motions are overlaid on the map obtained by the Plank mission of the area 
where the distribution of the warm dust and the orientation of the magnetic field are shown \citep{2016A&A...596A.103P}. 
The TTSs motion is very coherent as expected in a young association that still keeps memory of the motion of the parent molecular cloud; 
the young  gravitationally bound open cluster contains hundreds of members with similar ages and composition; however, the kinematical memory of this common origin will be lost over time due to the interaction with the large scale gravitational field of the Galaxy. As pointed out back in the early 90's  \citep{1992ApJ...395..501G}  the TTSs star move roughly perpendicularly
to the direction of the magnetic field traced by the polarization of the dust.

\begin{figure}
   \centering
   \includegraphics[width=9cm]{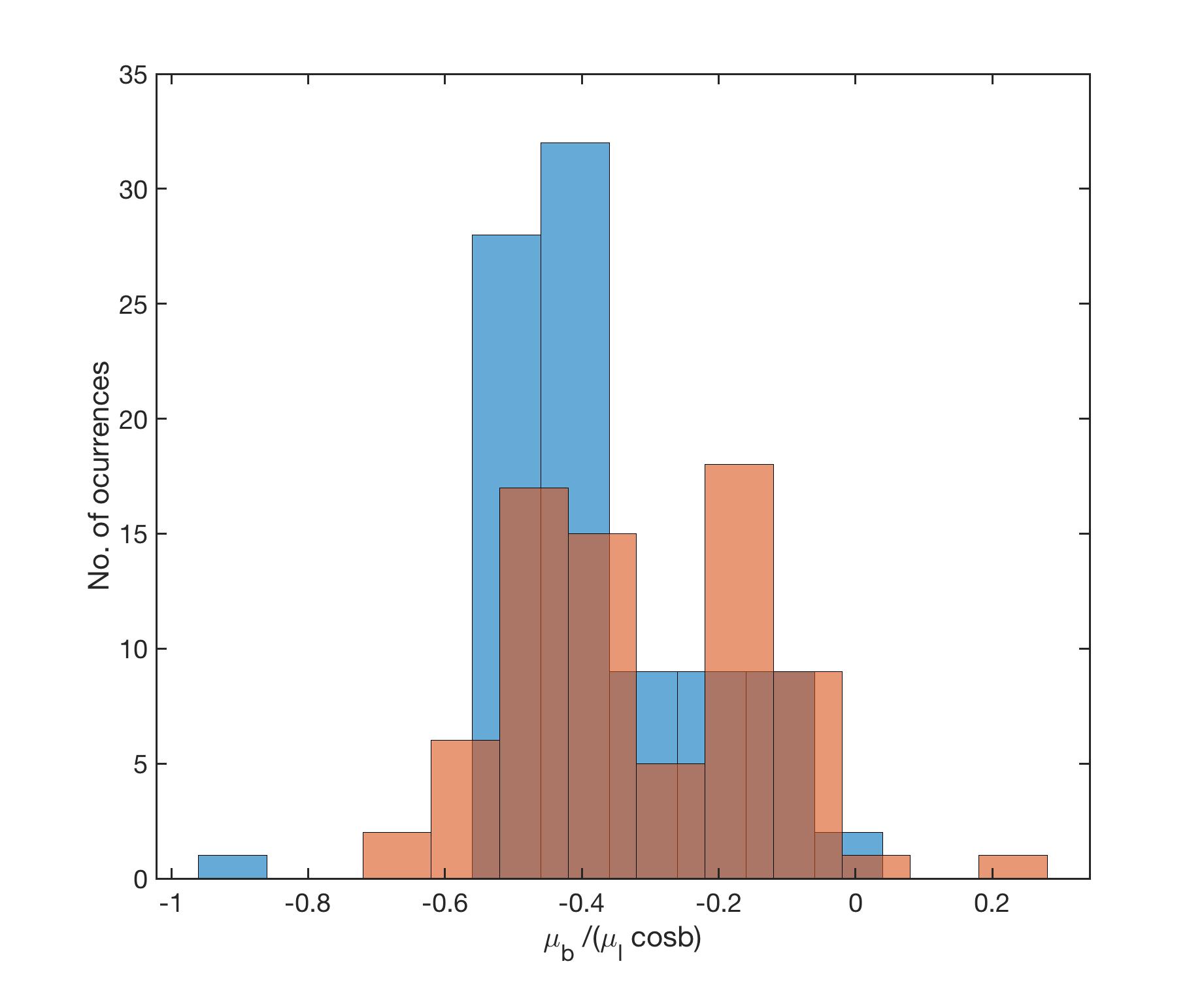}
      \caption{ Histogram of the distribution of the ratio $\mu_b/ \mu_l \cos(b)$ obtained for the known CTTSs  (blue) and WTTSs  (red)
      in the TMC.   }
         \label{fig5}
   \end{figure}

There are mild differences between WTTSs and CTTSs. This is best shown by 
the $\mu_b/\mu_l$ ratio (see Figs. \ref{fig4}, \ref{fig5}); while CTTSs have a single peak distribution
with a smooth tail to small ratios, the WTTSs display a double peaked distribution. As CTTSs are younger stars, it is expected that they are better
coupled to the gas and keep a common kinematics. This effect is also observed in the spatial distribution of the accreting TTSs and the WTTSs;
CTTSs are located  close to the molecular cloud while WTTSs are more sparsely distributed \citep{2010ApJS..186..111L}.
Two outliers are detected:  the CTTSs, LkH$\alpha$~332 ($2MASS J04420777+2523118$) and the 
WTTSs $2MASS J04555288+3006523$; LkH$\alpha$~332 is at 11.3 arcsec of V1000~Tau, another TTSs.

\subsubsection{k-cluster analysis}
   \begin{figure}[h!]
\begin{tabular}{r}
   \includegraphics[width=8cm]{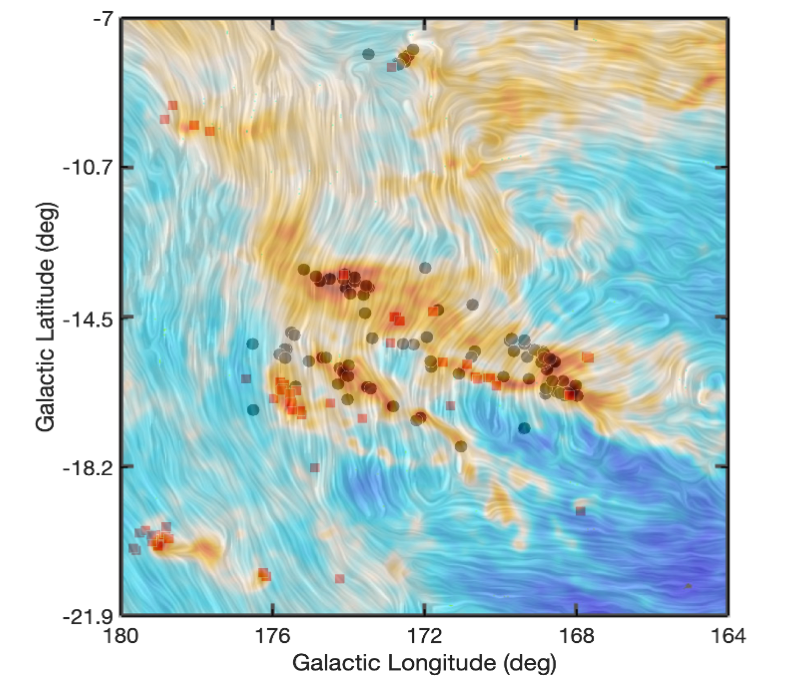} \\
      \includegraphics[width=7cm]{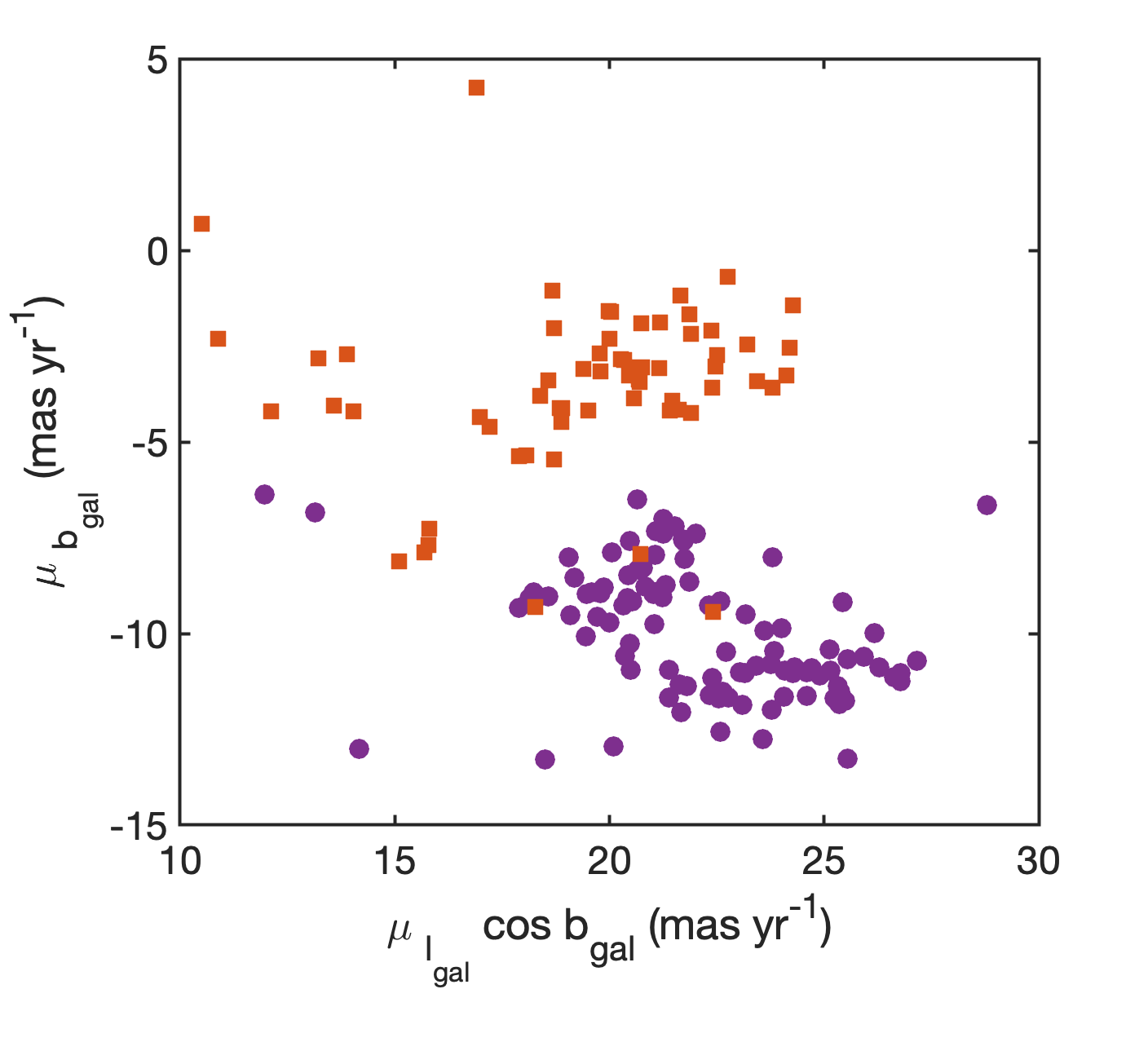} \\
      \includegraphics[width=7cm]{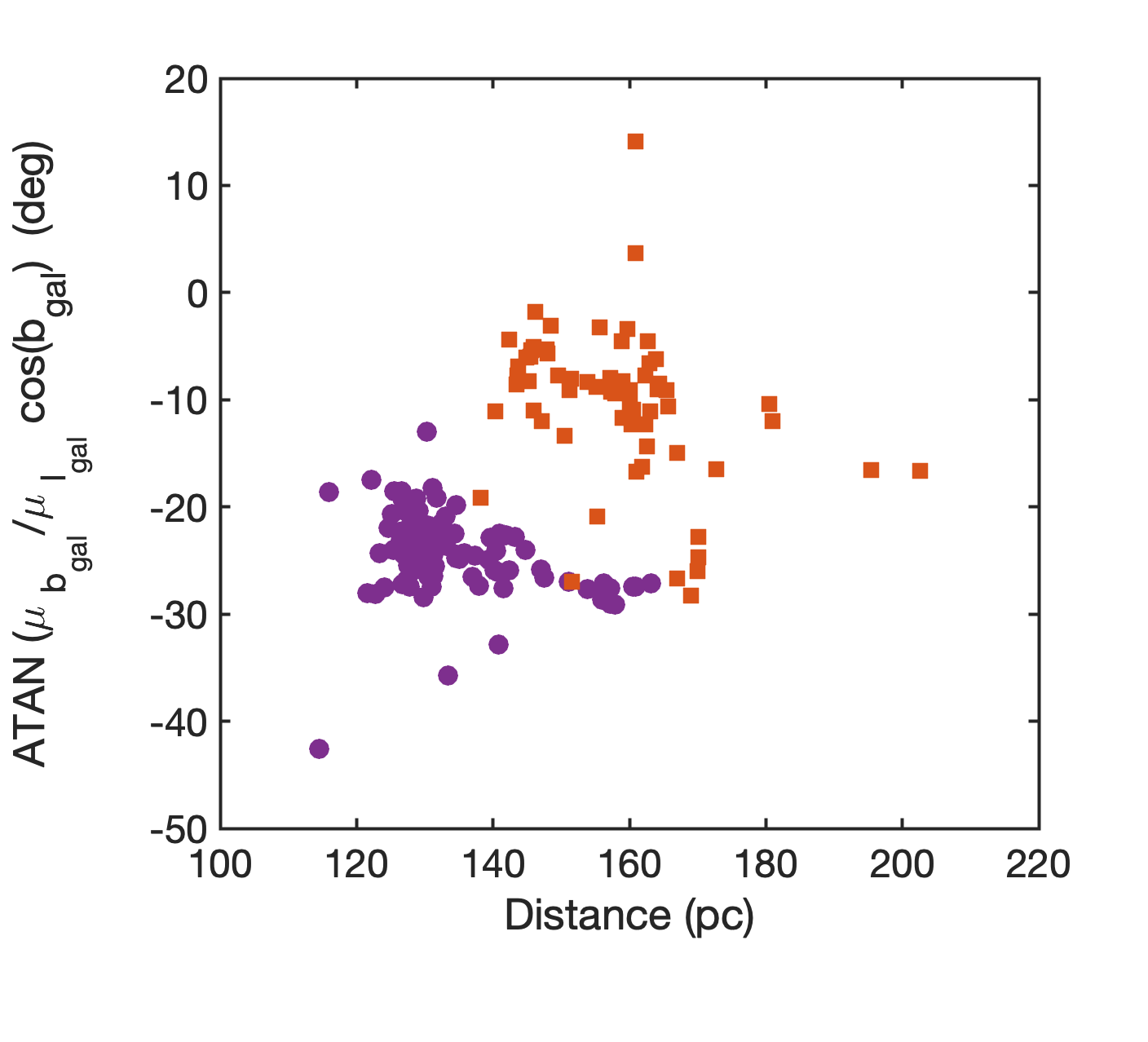} \\
      \end{tabular}
      \caption{ Location of the two groups identified by the decision tree  algorithm are represented. {\it Top}: sources location, {\it middle}: proper motions of the sources 
      in Galactic coordinates and {\it bottom:}  inclination of the proper motion with respect to the Galactic plane versus distance. The two groups are identified by orange
      squares (group 0) and  violet circles (group 1).}
         \label{fig6}
   \end{figure}
   
The TAMC occupies a large area in the sky and several kinematic groups have been identified within the association 
(see e.g. \citealt{2019A&A...630A.137G,2023AJ....165...37L}).  In this work, we have used the 
$k$-means++ clustering algorithm to determine which groups of stars in our qualification sample could be 
kinematically related. Given their young age, kinematically coherent sources may have been born out of the same gas 
cloudlet within the TAMC. This algorithm is an unsupervised machine-learning technique that performs a 
centroid-based analysis using iterative refinement \citep{10.1109/TIT.1982.1056489,kmeans++}. In our case, we 
carried out a multiparametric analysis that included both positional information (parallax) and kinematic information 
(the projections of the proper motion along the Galactic coordinates), i.e. our data set has three parameters. The 
$k$-means++ algorithm is not density-based like DBSCAN or OPTICS, but partitional so every object in the sample is 
assigned to one and only one cluster. This algorithm implicitly assumes that clusters are convex and isotropic and it 
performs best when applied to a mixture of Gaussian distributions with the same variances but perhaps different means. 
This mathematical precondition matches well the physical context of the TAMC sample when parallaxes and proper motions 
are considered. As our analysis is not including the coordinates of the sources but parallaxes and proper motions, the 
actual distributions of the data are rather isotropic and normally distributed. 

For each one of the 167 TTSs (92 CTTSs and 75 WTTSs),  a three-dimensional vector was generated. An initial set of 
$k$ points (centroids) was defined in this three-dimensional space and $k$ clusters were created by associating 
every vector with the nearest centroid. In a second step, a new centroid was computed for each cluster by 
calculating the mean among the vectors. Later, the association of each vector to a given cluster was re-evaluated
for the new centroid. The process was repeated until convergence was reached.

Here, we used the k-means++ algorithm as implemented by
the Python library Scikit-learn \citep{scikit-learn}.  Before applying the $k$-means++ algorithm, we scaled the data set using $Z$-score 
normalization: found the mean and standard deviation for each set of variables used in the analysis, subtracted the 
relevant mean from each value, and then divided by its corresponding standard deviation. Although the variables 
may originally have different variances, $Z$-score normalization standardizes variances and avoids placing more 
weight on variables with smaller variance when applying $k$-means++. Distance assignment between the data points in 
our sample assumed a Euclidean metric (other metrics gave consistent results) and we used the elbow method to 
determine the optimal value of clusters, $k$, that minimized the sum of the distances of all data points to their
respective cluster centres. Test analyses using OPTICS on the same data set led to similar results. 

This procedure was applied to the list of bona-fide TTSs (Table 2), but in order to take into account the 
associated uncertainties and their correlations, we generated 10$^{4}$ instances of the sample as explained in 
Appendix \ref{appendix_RFM} and for each synthetic instance, we determined the clusters through the $k$-means++ 
algorithm. In almost all the cases, our clustering analysis produced two statistically significant clusters
which are represented in Fig. 6. Sources in the first group are located
at an average distance of 160~pc while those in the second group are found at 130~pc. 
This clean classification in two groups differs from the results of the previous works by \citet{2019A&A...630A.137G} and
\citet{2023AJ....165...37L}. \citet{2019A&A...630A.137G} uses Hierarchical Association Clustering Mode (HACM)
which is a non-parametric statistical approach used for clustering analysis. HACM finds  the modes of
a kernel-based estimate of the density of points in the working space and groups the data points associated with the same modes
into one cluster with arbitrary shape. Clustering by mode identification requires only the space and the bandwidth of the kernel to be defined.
In particular, \citet{2019A&A...630A.137G}  make use of a 5-dimensional space consisting in the equatorial coordinates of  the sources, 
their proper motions in equatorial coordinates and the parallax $(\alpha, \delta, \mu_{\delta}, \mu_{\alpha}\cos\delta, \pi)$. 
The algorithm identifies 21 clusters hierarchically linked in the dendrogram only 4 of them having more than 10 sources. This result is natural 
since the coordinates are taken into account and the stellar population is sparsely distributed in the TAMC thus, stars with similar proper motions and parallaxes
located in different areas are identified as independent groups.  \citet{2023AJ....165...37L} follows a different approach. 
\textit{Gaia} astrometry is analyzed in terms of {\it proper motions offsets} defined as the differences between the observed proper motion of the star and the
the motion expected at the celestial coordinates and parallactic distance of the star for a specified
space velocity. This procedure is set to minimize the projection
effects given the large extent of the TAMC and the offsets are calculated with respect to 
the expected LSR velocity given in term of the velocity vector (U,V,W) = (-16, -12, -9)
km s$^{-1}$ , which according to \citet{2018AJ....156..271L} approximates the median velocity of Taurus
members. 9 groups are identified in this manner (see Fig. 1 and Table~6 in  \citet{2023AJ....165...37L} ). 
These groups do not coincide with those in \citet{2019A&A...630A.137G} but some of the levels in the
hierarchy defined by  \citet{2019A&A...630A.137G}  contain some of these groups.
In our work, the classification is made with a $k$-means++ clustering algorithm which does not provide a
hierarchical classification, neither a dendrogram suitable for comparison with  \citet{2019A&A...630A.137G}.
Moreover, the analysis in Sect. 3.1, as well as previous works \citep{1992ApJ...395..501G}, show that 
galactic coordinates are better suited for the kinematical analysis of the region (see Fig. 4). Proper motions are converted 
from equatorial to galactic coordinates and the cluster analysis is carried in a 3-dimensional space including these proper motions and the parallax
$(\mu_{b},\mu_{l}cosb, \pi)$. The number of groups identified by the algorithm is reduced to 2 (groups 0 and 1) with a clear difference between them
in the proper motion space and distance.  Groups L1544 and L1517 in \citet{2023AJ....165...37L} belong to group 0 and 1 in this work, respectively.
However, the many groups identified  by  \citet{2023AJ....165...37L}  in the center of the  TAMC belong to one of the two
underlying groups identified in this work. Our method does not assume a common LSR velocity for the full TAMC, as
in  \citet{2023AJ....165...37L} and does nor require to know the radial velocity of the stars to evaluate the differences in the
(U,V,W) because there very few sources with accurate determinations of the radial velocity (see, \citet{2019A&A...630A.137G} ).
We just use high quality data from \textit{Gaia} DR3 and the analysis of the properties of the TAMC to use optimal dynamical tracers for the classification.
This strategy shows to be very efficient in the description of the region.

\subsection{Kinematics of the TTS candidates and membership probability}

So far, we have discussed the kinematics in the
TAMC based on the sample of 167 'bona fide' TTS. 
In this section, we use
the kinematic classification
performed in Sec. 3.1 in order to assign a membership
probability to each group of each of the 13 TTS candidates and to that purpose,
we build a logistic regression model using the Scikit-learn
package \citep{scikit-learn}. A detailed explanation of this method
with an application to astrophysics is provided in \citet{2018ExA....45..379B};
in this work, we only provide a simple explanation with the basics of the
methodology as a guide to the reader for our particular case.\par

We work with a qualification sample of $N = 167$ stars (the 'bona fide' TTS)
that are classified into two groups based on their values of
$(\mu_{l^{*}}, \mu_{b}, \pi)$ as explained in the
previous section, so the classification depends
on three features and is
binary (groups $k = \{1, 0\}$). In this case, the logistic regression model
gives the probability that a given star $x_{i}$ belongs to Group 1, since
the probability of belonging to Group 0 is the complementary:

\begin{equation}
 p(x_{i} \in 1) = \frac{1}{1 + \exp{-(\beta_{0} + \mu_{l^{*}} \beta_{1} + \mu_{b} \beta_{2} + \pi \beta_{3})}}
\end{equation}

\noindent
where the $\beta_{m}$ are the logistic regression parameters
fitted using the qualification sample, that have a median value
of $\beta_{0} = 17.052$, $\beta_{1} = -0.085$, 
$\beta_{2} = 0.616$, $\beta_{3} = -1.575$ (see Appendix \ref{appendix_LBA} for 
more details on how these parameters were computed taking into 
account the associated errors on proper motions and parallax).
We analyzed the confusion matrix and only one star is misclassified, so
this choice of parameters (proper motions in galactic coordinates and
parallax) gives a good separation of the groups.
 \par

Once the logistic regression model is trained, we apply it
to the sample of TTS candidates, and following the methodology
explained in Appendix \ref{appendix_LBA}, we have taken into account the errors
in the measurements and computed a median probability of belonging
to any of the two groups; these probabilities are shown in Table \ref{table:3}. 
The location of the candidates in the proper motions-distance diagram is shown in Fig.~7. 
Only sources No. 3 (J04423721$+$3401492) and No. 12 (V600~Aur) display properties very different from those of the qualification sample.
We want to highlight that
membership probability has been computed
even for those sources with astrometric measurements 
of moderate quality: J04290082$+$3152597 (RUWE: 1.413), 
J05000310$+$3001074 (RUWE: 1.615) and J05005485$+$3229168 (RUWE: 1.773) and 
even bad quality  J04455129$+$1555496 (RUWE: 4.6815).

      \begin{table*}[h]
   \caption{Classification of the 13 TTs candidates
   after applying the Logistic Regression model with variables
   ( $\pi$ , $\mu_{l_{gal}} \cos {b_{gal}}$, $\mu_{l_{gal}}$).}
   \label{table:3}
   \centering
      \footnotesize
   \begin{tabular}{ccccccc}
   \hline \hline
   No. & Name & RA & DEc & P(Group 0) & P(Group 1) & RUWE\\
   & 2MASS &        &            &            &          &\\
   \hline
   1 & J04090973+2901306  & 62.29073244819 & 29.0249285705 & 0.9974 & 0.0026 &  1.011 \\
2 & J04290082+3152597 & 67.25351322368 & 31.88312957994    & 0.9919 & 0.0081 &  1.413\\
3 & J04423721+3401492 & 70.65496319697 & 34.03025801994    & 0.1347 & 0.8653 & 1.482\\
4 & J04455129+1555496  & 71.46378376348 & 15.93038180969   & 0.0026 & 0.9974 & 4.685\\
5 & J04510713+1708468 & 72.77977787635 & 17.14620505952    & 0.6846 & 0.3154 & 1.122\\
6 & J04590305+3003004& 74.76271085992 & 30.04998394972     & 0.8744 & 0.1256 & 1.242\\
7 & J04595003+3049082   & 74.958677841 & 30.81864271002    & 1.0    & 0.0    & 23.043\\
8 & J05000310+3001074   & 75.01300900515 & 30.01899690108  & 0.9288 & 0.0712 & 1.615\\
9 & J05005485+3229168   & 75.22852314723 & 32.48790302855  & 0.7567 & 0.2433 & 1.773\\
10 & J05050770+2923477   & 76.28212032867 & 29.39648322479 & 0.9993 & 0.0007 &  1.04\\
11 & J05074697+3120186   & 76.94572807385 & 31.33837744853 & 0.8431 & 0.1569 &  1.211\\
12 & J05102234+3126402   & 77.59293165054 & 31.44435544281 & 1.0 & 0.0 &  3.529\\
13 & J05240794+2542438   & 81.03309628183 & 25.71205212925 & 0.8749 & 0.1251 & 1.368\\
\hline
   \end{tabular}
   \end{table*}

  \begin{figure}[h!]
   \includegraphics[width=9.5cm]{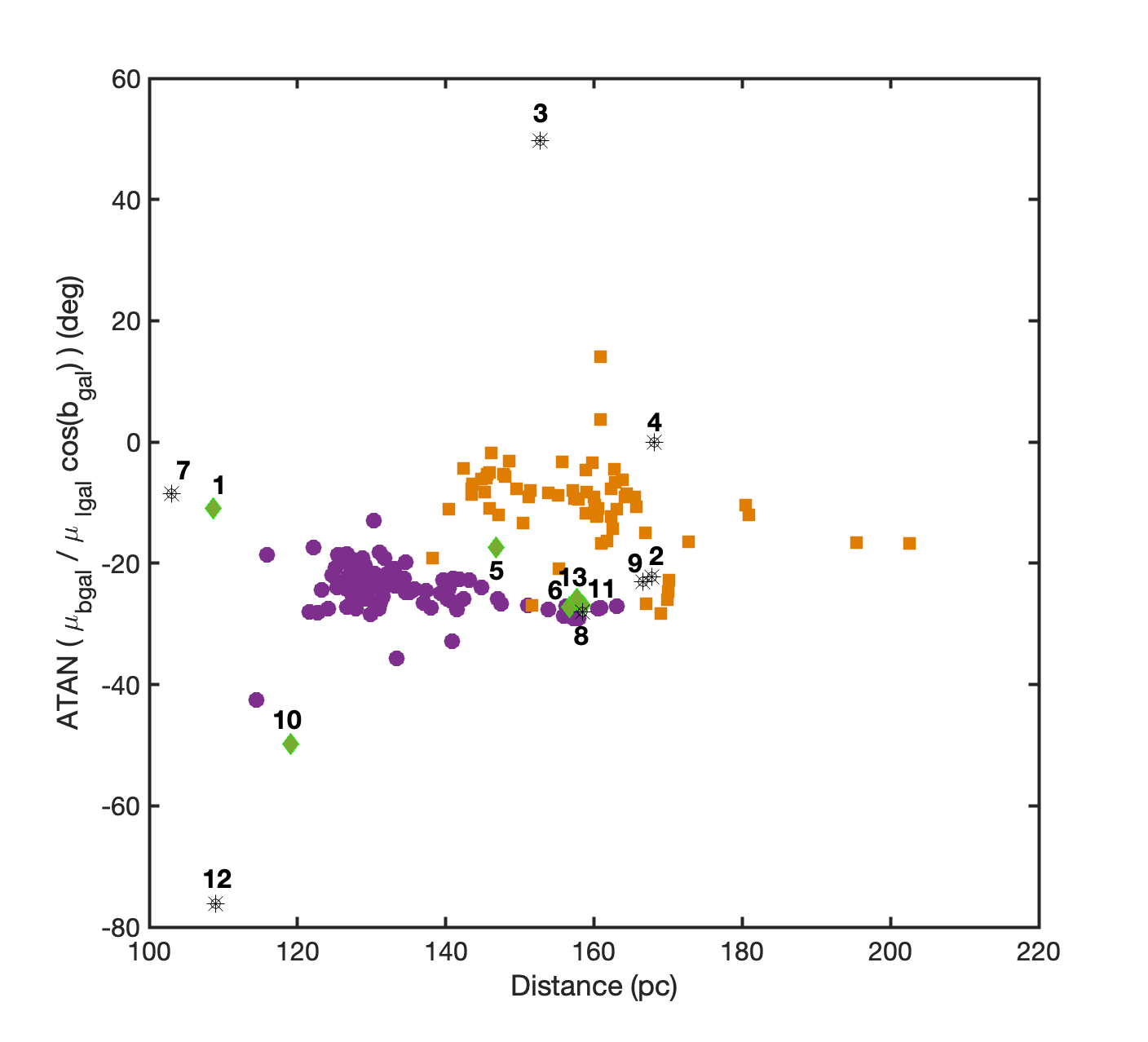} \\
      \caption{ As in the bottom panel of Fig. 6. The location of the 13 candidates to TTSs examined in this work is indicated and labeled
      according to their entry in Table 3. The candidates are marked according to their RUWE value: green diamonds (RUWE $\leq$ 1.4)
      and black dartboards (RUWE $>1.4$).
      }
         \label{fig6}
   \end{figure}

\section{Spectral information on the candidates}

In addition to the kinematical information, {\it Gaia} DR3 provides data on the 
effective temperature, surface gravity and metallicity  for seven of the thirteen
candidate sources  (see Table \ref{table:4}). These data are also indicated in the table
and compatible with the candidates being late type stars
as expected for the TAMC  stellar population. 

A peculiar source is HO Aur. The metallicity derived by {\it Gaia} for this source is very low (-3.9707) and given
the high quality of the measurements (RUWE $<$ 1.4), we feel inclined to rely on the data provided by the mission 
and tentatively identify HO Aur as a Population II field star.

\begin{figure}
   \centering
   \includegraphics[width=9cm]{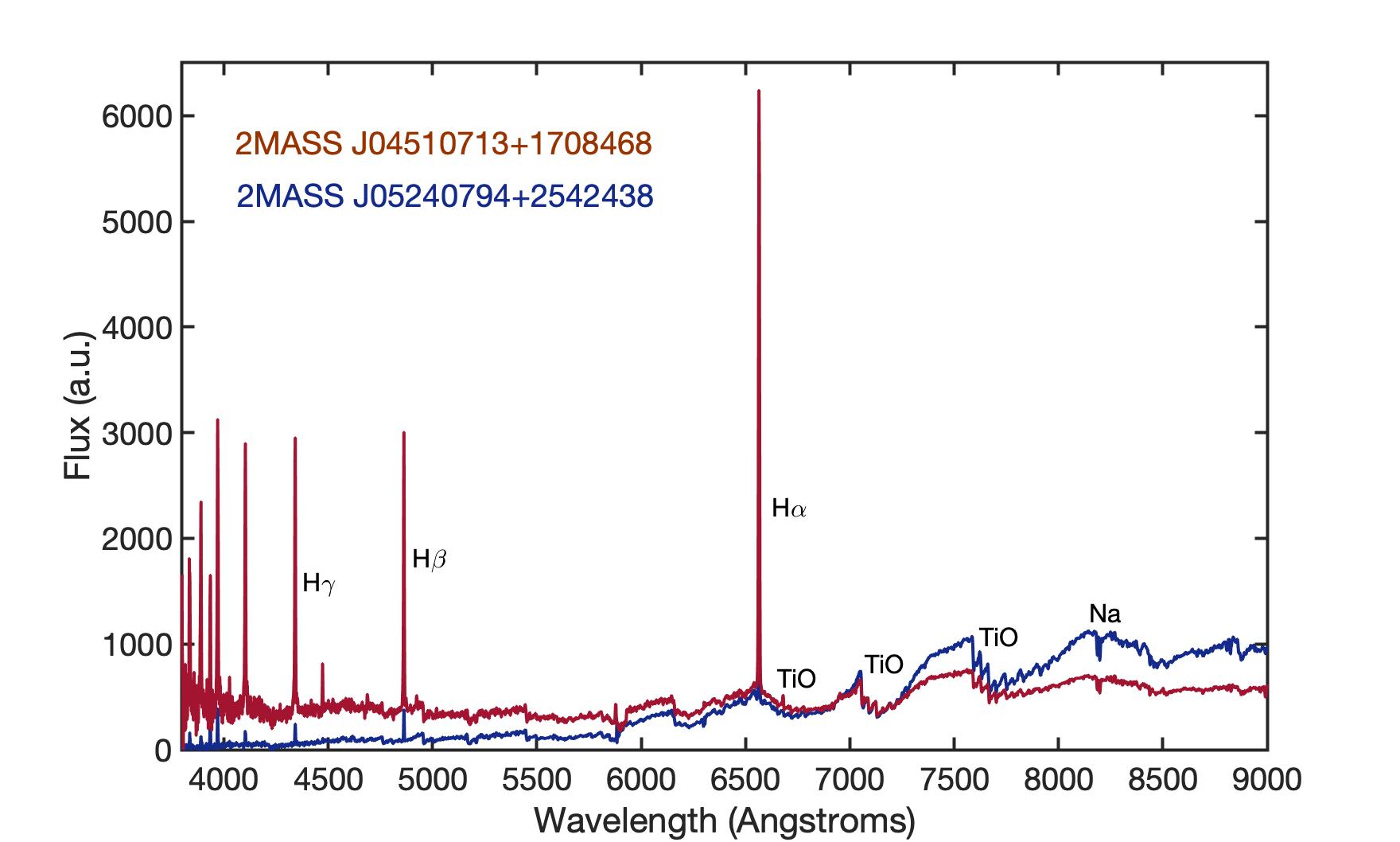} \\
   \includegraphics[width=9cm]{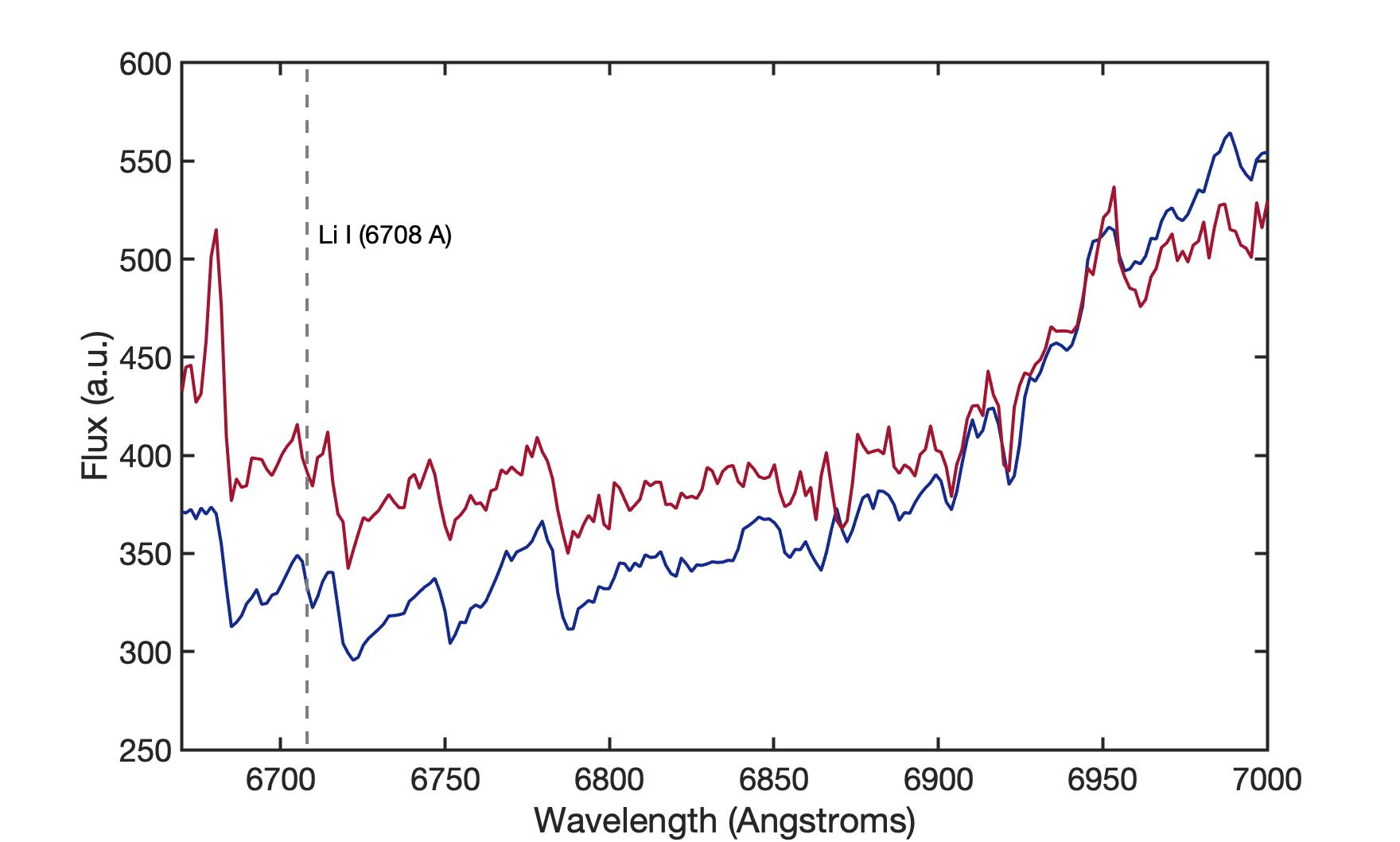}
      \caption{ LAMOST spectrum of the two newly identified accreting brown dwarfs in the TAMC. 
      {\it Top:} LAMOST spectrum; the Balmer series and the TiO bands are readily identified. {\it Bottom:} Zoom around the Li I feature. }
         \label{fig8}
   \end{figure}

LAMOST spectra are available for few sources and are compatible with the effective temperatures derived by \textit{Gaia}. 
In particular, the  spectra of $J04510713+1708468 $ and  $J05240794+2542438$ are compatible with late M spectral type 
and display strong TiO bands. We have downloaded the spectra from the Vizier server and measured
the TiO-5 index \citep{2002AJ....123.2828C} for these sources obtaining values of 1.28 and 0.85,  respectively
which confirms their  M-type classification. The spectra also display a strong  Na I absorption feature at 8183/8199~\AA\  (see Fig. \ref{fig8}). 
which is gravity sensitive \citep{2010A&A...517A..53M} unfortunately, the spectral resolution is  too low to cross-check the 
effective gravity determined by \textit{Gaia} for these sources. The spectra also show Li I absorption at 6708~\AA\
but again, equivalent widths are very uncertain and thus, it is difficult to provide an age estimate of the sources.

The most noticeable characteristics of the spectra of these stars are the prominent emission lines, in particular the Balmer series and the Ca II doublet
(see  Fig. \ref{fig8}).  This indicates that  the two M dwarfs are young and possibly accreting (see \citealt{2000ARA&A..38..485B}
and references therein). This is consistent with the method used by GdC2015 to search for TTSs candidates that was based on selecting sources that was based of the UV excess of the sources; 2MASS~$J04510713+1708468 $ has $FUV-NUV=1$ and $J-K=0.9$ and 2MASS~$J05240794+2542438$ has $FUV-NUV=1.4$ 
and $J-K=0.97$.

HD 281691 is a G8 star located in front of the TAMC, at a distance of 108.7 pc, according to the {\it Gaia} parallax and 
its proper motions are significantly different from those of the TTSs in the system. HD 281691 has a nearby companion
located at projected distance of 6.78 arcsec (738 au) \citep{2013ApJ...773...73J} and it is often included in surveys for 
debris disk   \citep{2017AJ....154..245M} . The infrared excess from the disk and the UV emission from  the active star resulted 
in its detection by the GdC2015 survey. 

 HD 30171 (J04455129$+$1555496) was detected as a TTS candidate in the GdC2015 survey and it is known to be a WTTS since the 
early searches for Li I absorption of the X-ray sources detected by the ROSAT satellite in TAMC \citep{2000A&A...359..181W}. 
As such, it is included in the catalogue used for qualification of the TTSs sample \citep{2017A&A...599A..14J} and it is a
clear member of Group 2.

V600 Aur is an active X-ray source which was included in the sample studied by \citet{2010ApJ...723.1542X} for the determination of the 
Li I equivalent width (see Table 1 in the article, [LH98] 173). Its rapid rotation period (2.201 days) and high Li I abundance 
support its pre-main sequence nature and TAMC membership. Note that though its kinematics deviates apparently from that of the TAMC members,
this may be caused by the low quality of the \textit{Gaia} measurements (RUWE$=$3.529). 

J04590305$+$3003004 and J05000310$+$3001074 are included in the {\it Catalogue of Stars in the Northern Milky Way Having H$\alpha$  in Emission}
 \citep{1999A&AS..134..255K} and display infrared excess emission \citep{2023AJ....165...37L}.
Recently, J04590305$+$3003004  has  been reported as a 8.8 Myr old dipper star, a subgroup among the  young stellar objects 
that exhibit dimming variability in their light curves (drops in brightness by 10\%$-$50\%), attributed to the occultation of the star by the 
circumstellar disk material \citep{2022ApJS..263...14C}.

Little or no information is available about J04290082$+$3152597,  J04423721$+$3401492, J04595003$+$3049082, J05005485$+$3229168 and
J05050770$+$2923477.  Two of them,  which according to our work are very good candidates to being members of the sparse young stellar population
of the TAMC however, J04423721$+$3401492 kinematics differs significantly from that of the TAMC members and the quality of the \textit{Gaia} 
measurements is quite good (RUWE$=$1.482).

In summary, 3 out the 13 candidates identified in the GdC2015 survey are known pre main sequence stars namely, 
HD 30171, V600 Aur and  J04590305$+$3003004. Another two are not TTSs associated to the TAMC;  HD 281691 is a field G8-type star 
located in front of the TAMC with a nearby companion at a projected distance of 738 AU and HO~Aur has a too low metallicity
to belong to the TAMC. The rest (excluding J04423721$+$3401492) are very reliable candidates and two of them, in particular, have been confirmed as new late 
type TTSs from their LAMOST spectrum:  $J04510713+1708468 $ and  $J05240794+2542438$.

    \begin{table*}
   \caption{Stellar properties}
   \label{table:4}
   \centering
      \footnotesize
   \begin{tabular}{cccccc}
   \hline \hline
   Name & Teff & log(g)   & [Fe]/[H] ratio & LAMOST spectrum & Comments \\
   2MASS & (K)       & [cm$^2$/s]    &            &            &       \\
   \hline
J04090973+2901306 & 5035 & 4.3171 & -0.5184 & Y & HD 281691, Debris Disk   \\
J04290082+3152597 & 4735.8 &4.5392 & -0.1402 &N& \\
J04423721+3401492 &   &  &  & N & \\
J04455129+1555496 &   &  &  & N & HD 30171\\
J04510713+1708468 & 3385.4 & 4.2684 & 0.1421 & Y &Late type dwarf with Emission Lines \\
J04590305+3003004 & &   &    & N & \\
J04595003+3049082 & &   &    & N& \\
J05000310+3001074 & &   &   & N&  GZ Aur\\
J05005485+3229168 & &   &    & N& \\ 
J05050770+2923477 & 4133.7 &   4.4843 & 0.2099 & N & \\
J05074697+3120186 & 5718.9 & 4.1804 & -3.9707 & N& HO Aur, Population II\\
J05102234+3126402 & 4899.5 & 4.3814 & -0.5804 & Y & V600 Aur, LiI (Xing 2010)\\
J05240794+2542438 & 3529.2 & 4.044 &0.1507 & Y &ate type dwarf  with Emission Lines\\

\hline
   \end{tabular}
   \end{table*}

\section{Discussion: dynamical coupling between the TAMC and the TTSs in the region}

The radial velocity of the TTSs also provides  fundamental information to determine the coupling between the molecular gas in the TAMC and the young stellar population (see {\it e.g.} \citealt{1992ApJ...395..501G}). This determination is relevant for two purposes. Firstly, it indicates the degree of dynamical relaxation of the young stellar association.
Also, if the coupling is strong, this information can be used to estimate the motion of the molecular gas in the plane of the sky which otherwise, is unfeasible;
if the dynamical coupling between the gas and the stars is strong (if the radial velocities are similar) then, the motion of the clouds in the plane of the sky
can be inferred from the stellar motion and the 3D velocity vector of the molecular gas can be restored. 

There have been several attempts to examine this coupling, in particular that by \citet{2019A&A...630A.137G} based on \textit{Gaia} DR2 data. 
In their approach, they selected several TTSs (28) and compared their radial velocity with that of the molecular gas in the nearest area within the complex
(see Fig. 15 in \citealt{2019A&A...630A.137G}). There is a well known velocity gradient in the TAMC which is observed in the $^{12}$CO and $^{13}$CO maps \citep{2008ApJ...680..428G, 2008ApJS..177..341N};
the radial (V$_{LSR}$) velocity decreases from 8.5 to 5.5 km~s$^{-1}$ roughly as Galactic longitude increases (see Fig. 7 to 16 in \citealt{2008ApJS..177..341N}).
\citet{2019A&A...630A.137G} attempted to search traces of this gradient in the stellar radial velocity with uncertain results. In their work, they selected the 
TTSs with good radial velocity measurements from \textit{Gaia} DR2 and extracted from the $^{13}$CO data cubes the velocity of the nearby (in projection) 
molecular gas (the beam size of the radio maps is 45" and the Nyquist-samplex pixel $\sim$20"). They found that stars near L1495, B213 and B216 
(which are at $l_{gal} ~\sim 170^o$) have larger $V_{LSR}$ than stars at L1536 ($l_{gal} ~\sim 175^o$) but the correlation is not good further than these two extreme groups.   
Moreover, the uncertainties of the $^{13}$CO $V_{LSR}$ measurements are high given the large broadening of the $^{13}$CO profiles and the extent of the 
area within the beam.  In summary, though the attempt is to be commended the results are non-conclusive.

\begin{figure}
   \centering
   \includegraphics[width=9cm]{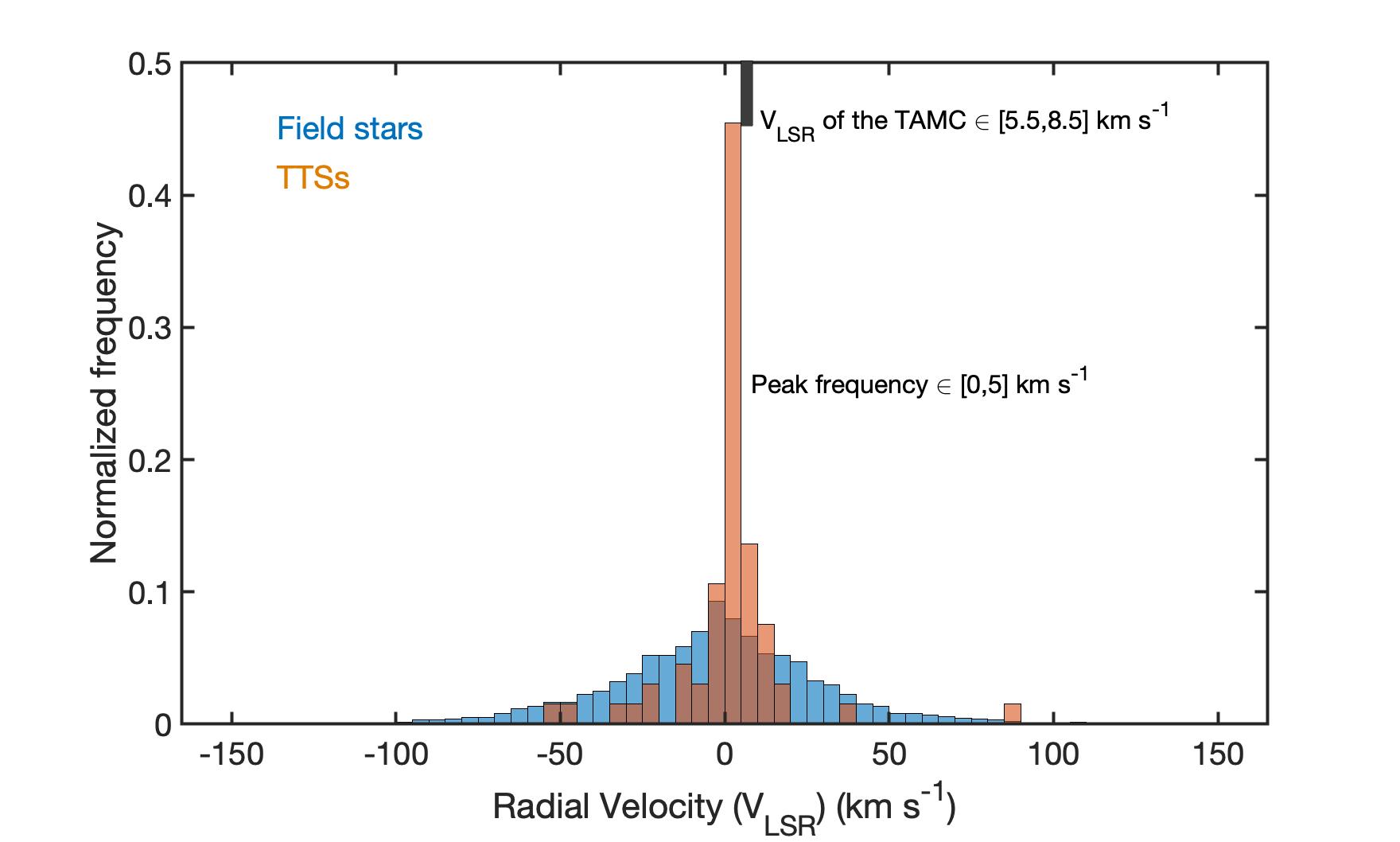} \\
      \caption{ Histogram of the frequencies of the distribution of stars in the TAMC field in terms of V$_{LSR}$. 
      The distribution of field stars is compared with that of the {\it bona fide} TTSs. }
         \label{fig9}
   \end{figure}

A simpler, statistical approach may help to examine the stars-gas coupling. There are about 6.2 million stars with galactic coordinates ($l_{gal} \in [164^o, 185^o], b_{gal} \in [-24^o, -4^o]$) in the \textit{Gaia} DR3 survey. This number reduces to 697,990,  if only sources with accurate radial velocity (relative error $< 10$\%) are included. Among them, only 20,806 are located at a distance compatible with TAMC membership. We have evaluated the 
radial velocity distribution ($V_{LSR}$) for these field stars and compared it with that obtained for the TTSs with high quality
\textit{Gaia} DR3 radial velocity measurements (66 TTSs  of those 40 are  WTTSs and  26 are  CTTSs).
The velocity distribution of the field stars in the area is very broad with a slight
asymmetry towards bluewards shifted velocities however, the velocity distribution of the TTSs (both WTTSs and CTTSs) peaks at redwardshifted 
velocities of $\sim$ 2.5 km~s$^{-1}$ (see Fig. \ref{fig9}). The Kolmogorov-Smirnov test rejects the null hypothesis (both distributions are the same); the 
$p$-value is 1.32$\times 10^{-12}$.  Thus, as otherwise expected, the TTSs differ from field stars also in radial velocity.
The relevant issue for the sake of this discussion is that the peak frequency of the TTSs velocity is smaller than the V$_{LSR}$ of the TAMC 
thus, the TTSs as a whole seem to be already dynamically decoupled from the cloud.

\begin{figure}
\centering
 \includegraphics[width=9cm]{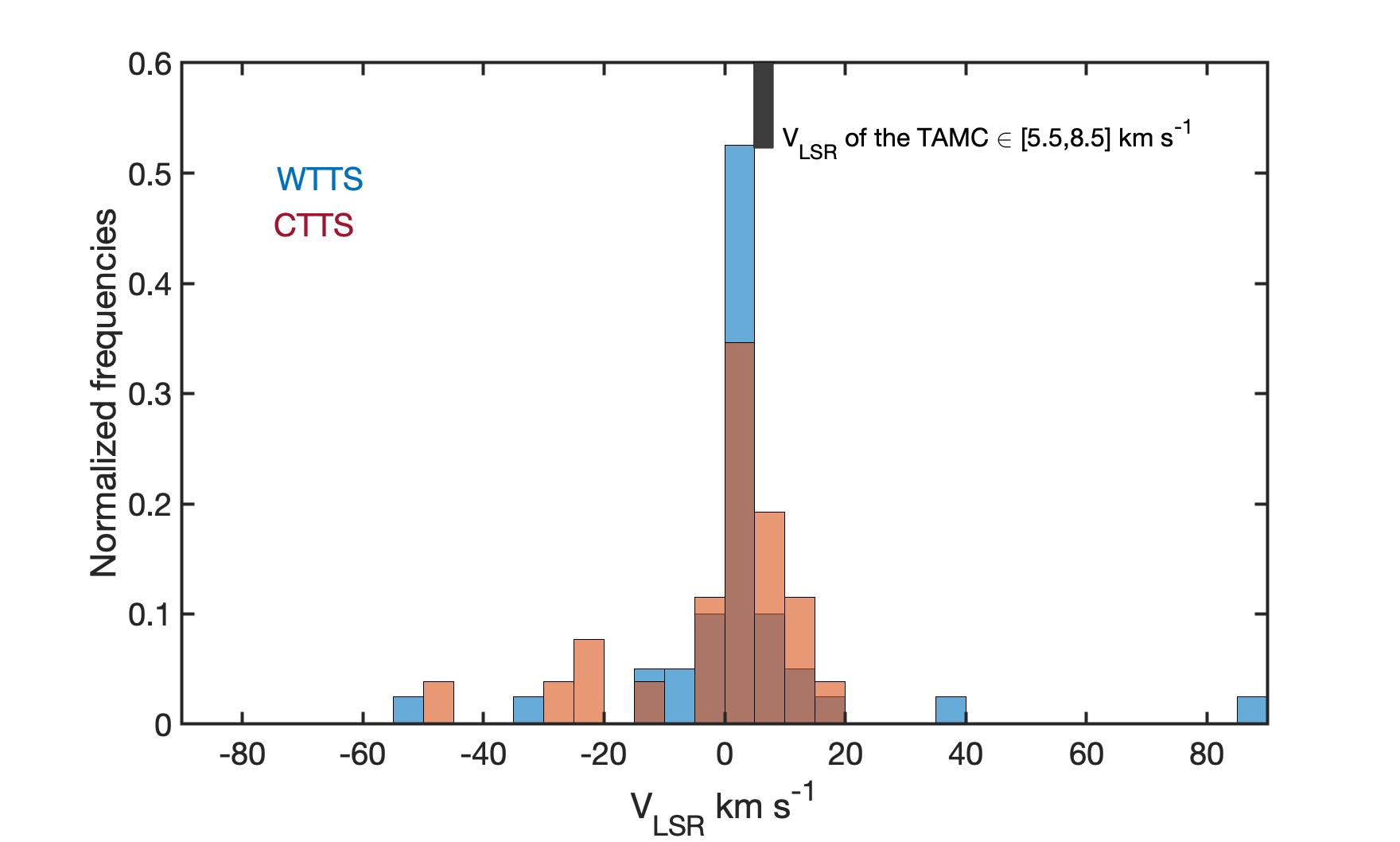} \\
 \caption{ Histogram of the frequencies of the distribution of TTSs stars in terms of V$_{LSR}$. 
 The distribution of the CTTSs  is compared with that of the WTTSs. }
  \label{fig10}
 \end{figure}

Unfortunately, the statistic is too poor to seek for differences between CTTSs and WTTSs. In spite of the apparent slight bias of the CTTSs 
towards a higher $V_{LSR}$ hinted  at in Fig. \ref{fig10}, the Kolmogorov-Smirnov test clearly indicates that both CTTSs and WTTS distributions are 
independent samples of the same underlying distribution with $p$-value 0.998.

\section{In summary}

In this work, \textit{Gaia} DR3 astrometric data have been used to review the list of candidates to TTSs selected by GdC2015 on the basis of their UV and IR colours. 
From the initial compilation of 63 sources, only 10 (16\%) have been found to be reliable members (shared kinematics and location
with well known members of the association). Most of these sources are located at low galactic latitudes, in the Auriga-Perseus area.
\textit{Gaia} spectroscopic information has shown that one of candidates pinpointed in the GdC2015 list, HO Aur, is a very low metallicity, 
population II star which cannot be member of this young association. All the rest of the sources are very likely TTSs; in particular, 
two of them have been identified as accreting late-type dwarfs from their LAMOST spectra: 2MASS J04510713+1708468 and 2MASS J05240794+2542438. 

This work has also produced some additional results. Firstly, a clean, reliable list of TTSs for astrometric studies has been produced. In addition,
the analysis of the radial velocity ($V_{LSR}$) of the TTSs in comparison to those of the field stars and the molecular cloud has shown that the association
has  clearly distinct kinematics and that the TTSs are no longer coupled to the molecular gas in the TAMC. As the sample is dominated by WTTSs,
this suggests that the decoupling is already significant  $10 - 100$~Myr after star formation begun.

\begin{acknowledgements}

This work has been partially financed by the Ministry of Science and Innovation through grants: ESP2017-87813-R and PID2020-116726RB-I00.
This research has made use of the SIMBAD database, operated at CDS, Strasbourg, France and the computational facilities of the Joint Center for Ultraviolet Astronomy in the campus of the Universidad Complutense de Madrid.
L.B.-A. acknowledges the receipt of a Margarita Salas postdoctoral
fellowship from Universidad Complutense de Madrid (CT31/21), funded by
the 'Ministerio de Universidades' with Next Generation EU funds.
This work has made use of data from the European Space Agency (ESA) mission
{\it Gaia} (\url{https://www.cosmos.esa.int/gaia}), processed by the {\it Gaia}
Data Processing and Analysis Consortium (DPAC,
\url{https://www.cosmos.esa.int/web/gaia/dpac/consortium}). Funding for the DPAC
has been provided by national institutions, in particular the institutions
participating in the {\it Gaia} Multilateral Agreement.
 \end{acknowledgements}

%
\bibliographystyle{aa} 
\bibliography{references} 

\begin{thebibliography}{37}
\expandafter\ifx\csname natexlab\endcsname\relax\def\natexlab#1{#1}\fi

\bibitem[{{Arthur} \& {Vassilvitskii}(2007)}]{kmeans++}
{Arthur}, D. \& {Vassilvitskii}, S. 2007, in Proceedings of the Eighteenth
  Annual ACM-SIAM Symposium on Discrete Algorithms, Society for Industrial and
  Applied Mathematics, Philadelphia, ed. G.~H. (Society for Industrial and
  Applied Mathematics), 1027--1035

\bibitem[{{Basri}(2000)}]{2000ARA&A..38..485B}
{Basri}, G. 2000, \araa, 38, 485

\bibitem[{{Beitia-Antero} {et~al.}(2018){Beitia-Antero}, {Y{\'a}{\~n}ez}, \&
  {de Castro}}]{2018ExA....45..379B}
{Beitia-Antero}, L., {Y{\'a}{\~n}ez}, J., \& {de Castro}, A. I.~G. 2018,
  Experimental Astronomy, 45, 379

\bibitem[{{Cantat-Gaudin} \& {Anders}(2020)}]{2020A&A...633A..99C}
{Cantat-Gaudin}, T. \& {Anders}, F. 2020, \aap, 633, A99

\bibitem[{{Capistrant} {et~al.}(2022){Capistrant}, {Soares-Furtado},
  {Vanderburg}, {Kounkel}, {Rappaport}, {Omohundro}, {Powell}, {Gagliano},
  {Jacobs}, {Kostov}, {Kristiansen}, {LaCourse}, {Schmitt}, {Schwengeler}, \&
  {Terentev}}]{2022ApJS..263...14C}
{Capistrant}, B.~K., {Soares-Furtado}, M., {Vanderburg}, A., {et~al.} 2022,
  \apjs, 263, 14

\bibitem[{{Cruz} \& {Reid}(2002)}]{2002AJ....123.2828C}
{Cruz}, K.~L. \& {Reid}, I.~N. 2002, \aj, 123, 2828

\bibitem[{{Gaia Collaboration} {et~al.}(2018){Gaia Collaboration}, {Brown},
  {Vallenari}, {Prusti}, {de Bruijne}, {Babusiaux}, {Bailer-Jones}, {Biermann},
  {Evans}, {Eyer}, {Jansen}, {Jordi}, {Klioner}, {Lammers}, {Lindegren},
  {Luri}, {Mignard}, {Panem}, {Pourbaix}, {Randich}, {Sartoretti}, {Siddiqui},
  {Soubiran}, {van Leeuwen}, {Walton}, {Arenou}, {Bastian}, {Cropper},
  {Drimmel}, {Katz}, {Lattanzi}, {Bakker}, {Cacciari}, {Casta{\~n}eda},
  {Chaoul}, {Cheek}, {De Angeli}, {Fabricius}, {Guerra}, {Holl}, {Masana},
  {Messineo}, {Mowlavi}, {Nienartowicz}, {Panuzzo}, {Portell}, {Riello},
  {Seabroke}, {Tanga}, {Th{\'e}venin}, {Gracia-Abril}, {Comoretto},
  {Garcia-Reinaldos}, {Teyssier}, {Altmann}, {Andrae}, {Audard},
  {Bellas-Velidis}, {Benson}, {Berthier}, {Blomme}, {Burgess}, {Busso},
  {Carry}, {Cellino}, {Clementini}, {Clotet}, {Creevey}, {Davidson}, {De
  Ridder}, {Delchambre}, {Dell'Oro}, {Ducourant},
  {Fern{\'a}ndez-Hern{\'a}ndez}, {Fouesneau}, {Fr{\'e}mat}, {Galluccio},
  {Garc{\'\i}a-Torres}, {Gonz{\'a}lez-N{\'u}{\~n}ez}, {Gonz{\'a}lez-Vidal},
  {Gosset}, {Guy}, {Halbwachs}, {Hambly}, {Harrison}, {Hern{\'a}ndez},
  {Hestroffer}, {Hodgkin}, {Hutton}, {Jasniewicz}, {Jean-Antoine-Piccolo},
  {Jordan}, {Korn}, {Krone-Martins}, {Lanzafame}, {Lebzelter}, {L{\"o}ffler},
  {Manteiga}, {Marrese}, {Mart{\'\i}n-Fleitas}, {Moitinho}, {Mora}, {Muinonen},
  {Osinde}, {Pancino}, {Pauwels}, {Petit}, {Recio-Blanco}, {Richards},
  {Rimoldini}, {Robin}, {Sarro}, {Siopis}, {Smith}, {Sozzetti}, {S{\"u}veges},
  {Torra}, {van Reeven}, {Abbas}, {Abreu Aramburu}, {Accart}, {Aerts},
  {Altavilla}, {{\'A}lvarez}, {Alvarez}, {Alves}, {Anderson}, {Andrei},
  {Anglada Varela}, {Antiche}, {Antoja}, {Arcay}, {Astraatmadja}, {Bach},
  {Baker}, {Balaguer-N{\'u}{\~n}ez}, {Balm}, {Barache}, {Barata}, {Barbato},
  {Barblan}, {Barklem}, {Barrado}, {Barros}, {Barstow}, {Bartholom{\'e}
  Mu{\~n}oz}, {Bassilana}, {Becciani}, {Bellazzini}, {Berihuete}, {Bertone},
  {Bianchi}, {Bienaym{\'e}}, {Blanco-Cuaresma}, {Boch}, {Boeche}, {Bombrun},
  {Borrachero}, {Bossini}, {Bouquillon}, {Bourda}, {Bragaglia}, {Bramante},
  {Breddels}, {Bressan}, {Brouillet}, {Br{\"u}semeister}, {Brugaletta},
  {Bucciarelli}, {Burlacu}, {Busonero}, {Butkevich}, {Buzzi}, {Caffau},
  {Cancelliere}, {Cannizzaro}, {Cantat-Gaudin}, {Carballo}, {Carlucci},
  {Carrasco}, {Casamiquela}, {Castellani}, {Castro-Ginard}, {Charlot},
  {Chemin}, {Chiavassa}, {Cocozza}, {Costigan}, {Cowell}, {Crifo}, {Crosta},
  {Crowley}, {Cuypers}, {Dafonte}, {Damerdji}, {Dapergolas}, {David}, {David},
  {de Laverny}, {De Luise}, {De March}, {de Martino}, {de Souza}, {de Torres},
  {Debosscher}, {del Pozo}, {Delbo}, {Delgado}, {Delgado}, {Di Matteo},
  {Diakite}, {Diener}, {Distefano}, {Dolding}, {Drazinos}, {Dur{\'a}n},
  {Edvardsson}, {Enke}, {Eriksson}, {Esquej}, {Eynard Bontemps}, {Fabre},
  {Fabrizio}, {Faigler}, {Falc{\~a}o}, {Farr{\`a}s Casas}, {Federici},
  {Fedorets}, {Fernique}, {Figueras}, {Filippi}, {Findeisen}, {Fonti},
  {Fraile}, {Fraser}, {Fr{\'e}zouls}, {Gai}, {Galleti}, {Garabato},
  {Garc{\'\i}a-Sedano}, {Garofalo}, {Garralda}, {Gavel}, {Gavras}, {Gerssen},
  {Geyer}, {Giacobbe}, {Gilmore}, {Girona}, {Giuffrida}, {Glass}, {Gomes},
  {Granvik}, {Gueguen}, {Guerrier}, {Guiraud}, {Guti{\'e}rrez-S{\'a}nchez},
  {Haigron}, {Hatzidimitriou}, {Hauser}, {Haywood}, {Heiter}, {Helmi}, {Heu},
  {Hilger}, {Hobbs}, {Hofmann}, {Holland}, {Huckle}, {Hypki}, {Icardi},
  {Jan{\ss}en}, {Jevardat de Fombelle}, {Jonker}, {Juh{\'a}sz}, {Julbe},
  {Karampelas}, {Kewley}, {Klar}, {Kochoska}, {Kohley}, {Kolenberg},
  {Kontizas}, {Kontizas}, {Koposov}, {Kordopatis}, {Kostrzewa-Rutkowska},
  {Koubsky}, {Lambert}, {Lanza}, {Lasne}, {Lavigne}, {Le Fustec}, {Le
  Poncin-Lafitte}, {Lebreton}, {Leccia}, {Leclerc}, {Lecoeur-Taibi},
  {Lenhardt}, {Leroux}, {Liao}, {Licata}, {Lindstr{\o}m}, {Lister}, {Livanou},
  {Lobel}, {L{\'o}pez}, {Managau}, {Mann}, {Mantelet}, {Marchal}, {Marchant},
  {Marconi}, {Marinoni}, {Marschalk{\'o}}, {Marshall}, {Martino}, {Marton},
  {Mary}, {Massari}, {Matijevi{\v{c}}}, {Mazeh}, {McMillan}, {Messina},
  {Michalik}, {Millar}, {Molina}, {Molinaro}, {Moln{\'a}r}, {Montegriffo},
  {Mor}, {Morbidelli}, {Morel}, {Morris}, {Mulone}, {Muraveva}, {Musella},
  {Nelemans}, {Nicastro}, {Noval}, {O'Mullane}, {Ord{\'e}novic},
  {Ord{\'o}{\~n}ez-Blanco}, {Osborne}, {Pagani}, {Pagano}, {Pailler},
  {Palacin}, {Palaversa}, {Panahi}, {Pawlak}, {Piersimoni}, {Pineau}, {Plachy},
  {Plum}, {Poggio}, {Poujoulet}, {Pr{\v{s}}a}, {Pulone}, {Racero}, {Ragaini},
  {Rambaux}, {Ramos-Lerate}, {Regibo}, {Reyl{\'e}}, {Riclet}, {Ripepi}, {Riva},
  {Rivard}, {Rixon}, {Roegiers}, {Roelens}, {Romero-G{\'o}mez}, {Rowell},
  {Royer}, {Ruiz-Dern}, {Sadowski}, {Sagrist{\`a} Sell{\'e}s}, {Sahlmann},
  {Salgado}, {Salguero}, {Sanna}, {Santana-Ros}, {Sarasso}, {Savietto},
  {Schultheis}, {Sciacca}, {Segol}, {Segovia}, {S{\'e}gransan}, {Shih},
  {Siltala}, {Silva}, {Smart}, {Smith}, {Solano}, {Solitro}, {Sordo}, {Soria
  Nieto}, {Souchay}, {Spagna}, {Spoto}, {Stampa}, {Steele},
  {Steidelm{\"u}ller}, {Stephenson}, {Stoev}, {Suess}, {Surdej}, {Szabados},
  {Szegedi-Elek}, {Tapiador}, {Taris}, {Tauran}, {Taylor}, {Teixeira},
  {Terrett}, {Teyssandier}, {Thuillot}, {Titarenko}, {Torra Clotet}, {Turon},
  {Ulla}, {Utrilla}, {Uzzi}, {Vaillant}, {Valentini}, {Valette}, {van Elteren},
  {Van Hemelryck}, {van Leeuwen}, {Vaschetto}, {Vecchiato}, {Veljanoski},
  {Viala}, {Vicente}, {Vogt}, {von Essen}, {Voss}, {Votruba}, {Voutsinas},
  {Walmsley}, {Weiler}, {Wertz}, {Wevers}, {Wyrzykowski}, {Yoldas},
  {{\v{Z}}erjal}, {Ziaeepour}, {Zorec}, {Zschocke}, {Zucker}, {Zurbach}, \&
  {Zwitter}}]{2018A&A...616A...1G}
{Gaia Collaboration}, {Brown}, A.~G.~A., {Vallenari}, A., {et~al.} 2018, \aap,
  616, A1

\bibitem[{{Gaia Collaboration} {et~al.}(2022){Gaia Collaboration}, {Vallenari},
  {Brown}, {Prusti}, {de Bruijne}, {Arenou}, {Babusiaux}, {Biermann},
  {Creevey}, {Ducourant}, {Evans}, {Eyer}, {Guerra}, {Hutton}, {Jordi},
  {Klioner}, {Lammers}, {Lindegren}, {Luri}, {Mignard}, {Panem}, {Pourbaix},
  {Randich}, {Sartoretti}, {Soubiran}, {Tanga}, {Walton}, {Bailer-Jones},
  {Bastian}, {Drimmel}, {Jansen}, {Katz}, {Lattanzi}, {van Leeuwen}, {Bakker},
  {Cacciari}, {Casta{\~n}eda}, {De Angeli}, {Fabricius}, {Fouesneau},
  {Fr{\'e}mat}, {Galluccio}, {Guerrier}, {Heiter}, {Masana}, {Messineo},
  {Mowlavi}, {Nicolas}, {Nienartowicz}, {Pailler}, {Panuzzo}, {Riclet}, {Roux},
  {Seabroke}, {Sordo{\o}rcit}, {Th{\'e}venin}, {Gracia-Abril}, {Portell},
  {Teyssier}, {Altmann}, {Andrae}, {Audard}, {Bellas-Velidis}, {Benson},
  {Berthier}, {Blomme}, {Burgess}, {Busonero}, {Busso}, {C{\'a}novas}, {Carry},
  {Cellino}, {Cheek}, {Clementini}, {Damerdji}, {Davidson}, {de Teodoro},
  {Nu{\~n}ez Campos}, {Delchambre}, {Dell'Oro}, {Esquej},
  {Fern{\'a}ndez-Hern{\'a}ndez}, {Fraile}, {Garabato}, {Garc{\'\i}a-Lario},
  {Gosset}, {Haigron}, {Halbwachs}, {Hambly}, {Harrison}, {Hern{\'a}ndez},
  {Hestroffer}, {Hodgkin}, {Holl}, {Jan{\ss}en}, {Jevardat de Fombelle},
  {Jordan}, {Krone-Martins}, {Lanzafame}, {L{\"o}ffler}, {Marchal}, {Marrese},
  {Moitinho}, {Muinonen}, {Osborne}, {Pancino}, {Pauwels}, {Recio-Blanco},
  {Reyl{\'e}}, {Riello}, {Rimoldini}, {Roegiers}, {Rybizki}, {Sarro}, {Siopis},
  {Smith}, {Sozzetti}, {Utrilla}, {van Leeuwen}, {Abbas}, {{\'A}brah{\'a}m},
  {Abreu Aramburu}, {Aerts}, {Aguado}, {Ajaj}, {Aldea-Montero}, {Altavilla},
  {{\'A}lvarez}, {Alves}, {Anders}, {Anderson}, {Anglada Varela}, {Antoja},
  {Baines}, {Baker}, {Balaguer-N{\'u}{\~n}ez}, {Balbinot}, {Balog}, {Barache},
  {Barbato}, {Barros}, {Barstow}, {Bartolom{\'e}}, {Bassilana}, {Bauchet},
  {Becciani}, {Bellazzini}, {Berihuete}, {Bernet}, {Bertone}, {Bianchi},
  {Binnenfeld}, {Blanco-Cuaresma}, {Blazere}, {Boch}, {Bombrun}, {Bossini},
  {Bouquillon}, {Bragaglia}, {Bramante}, {Breedt}, {Bressan}, {Brouillet},
  {Brugaletta}, {Bucciarelli}, {Burlacu}, {Butkevich}, {Buzzi}, {Caffau},
  {Cancelliere}, {Cantat-Gaudin}, {Carballo}, {Carlucci}, {Carnerero},
  {Carrasco}, {Casamiquela}, {Castellani}, {Castro-Ginard}, {Chaoul},
  {Charlot}, {Chemin}, {Chiaramida}, {Chiavassa}, {Chornay}, {Comoretto},
  {Contursi}, {Cooper}, {Cornez}, {Cowell}, {Crifo}, {Cropper}, {Crosta},
  {Crowley}, {Dafonte}, {Dapergolas}, {David}, {David}, {de Laverny}, {De
  Luise}, {De March}, {De Ridder}, {de Souza}, {de Torres}, {del Peloso}, {del
  Pozo}, {Delbo}, {Delgado}, {Delisle}, {Demouchy}, {Dharmawardena}, {Di
  Matteo}, {Diakite}, {Diener}, {Distefano}, {Dolding}, {Edvardsson}, {Enke},
  {Fabre}, {Fabrizio}, {Faigler}, {Fedorets}, {Fernique}, {Fienga}, {Figueras},
  {Fournier}, {Fouron}, {Fragkoudi}, {Gai}, {Garcia-Gutierrez},
  {Garcia-Reinaldos}, {Garc{\'\i}a-Torres}, {Garofalo}, {Gavel}, {Gavras},
  {Gerlach}, {Geyer}, {Giacobbe}, {Gilmore}, {Girona}, {Giuffrida}, {Gomel},
  {Gomez}, {Gonz{\'a}lez-N{\'u}{\~n}ez}, {Gonz{\'a}lez-Santamar{\'\i}a},
  {Gonz{\'a}lez-Vidal}, {Granvik}, {Guillout}, {Guiraud},
  {Guti{\'e}rrez-S{\'a}nchez}, {Guy}, {Hatzidimitriou}, {Hauser}, {Haywood},
  {Helmer}, {Helmi}, {Sarmiento}, {Hidalgo}, {Hilger}, {H{\l}adczuk}, {Hobbs},
  {Holland}, {Huckle}, {Jardine}, {Jasniewicz}, {Jean-Antoine Piccolo},
  {Jim{\'e}nez-Arranz}, {Jorissen}, {Juaristi Campillo}, {Julbe}, {Karbevska},
  {Kervella}, {Khanna}, {Kontizas}, {Kordopatis}, {Korn}, {K{\'o}sp{\'a}l},
  {Kostrzewa-Rutkowska}, {Kruszy{\'n}ska}, {Kun}, {Laizeau}, {Lambert},
  {Lanza}, {Lasne}, {Le Campion}, {Lebreton}, {Lebzelter}, {Leccia}, {Leclerc},
  {Lecoeur-Taibi}, {Liao}, {Licata}, {Lindstr{\o}m}, {Lister}, {Livanou},
  {Lobel}, {Lorca}, {Loup}, {Madrero Pardo}, {Magdaleno Romeo}, {Managau},
  {Mann}, {Manteiga}, {Marchant}, {Marconi}, {Marcos}, {Marcos Santos},
  {Mar{\'\i}n Pina}, {Marinoni}, {Marocco}, {Marshall}, {Polo},
  {Mart{\'\i}n-Fleitas}, {Marton}, {Mary}, {Masip}, {Massari},
  {Mastrobuono-Battisti}, {Mazeh}, {McMillan}, {Messina}, {Michalik}, {Millar},
  {Mints}, {Molina}, {Molinaro}, {Moln{\'a}r}, {Monari}, {Mongui{\'o}},
  {Montegriffo}, {Montero}, {Mor}, {Mora}, {Morbidelli}, {Morel}, {Morris},
  {Muraveva}, {Murphy}, {Musella}, {Nagy}, {Noval}, {Oca{\~n}a}, {Ogden},
  {Ordenovic}, {Osinde}, {Pagani}, {Pagano}, {Palaversa}, {Palicio},
  {Pallas-Quintela}, {Panahi}, {Payne-Wardenaar}, {Pe{\~n}alosa Esteller},
  {Penttil{\"a}}, {Pichon}, {Piersimoni}, {Pineau}, {Plachy}, {Plum}, {Poggio},
  {Pr{\v{s}}a}, {Pulone}, {Racero}, {Ragaini}, {Rainer}, {Raiteri}, {Rambaux},
  {Ramos}, {Ramos-Lerate}, {Re Fiorentin}, {Regibo}, {Richards}, {Rios Diaz},
  {Ripepi}, {Riva}, {Rix}, {Rixon}, {Robichon}, {Robin}, {Robin}, {Roelens},
  {Rogues}, {Rohrbasser}, {Romero-G{\'o}mez}, {Rowell}, {Royer}, {Ruz Mieres},
  {Rybicki}, {Sadowski}, {S{\'a}ez N{\'u}{\~n}ez}, {Sagrist{\`a} Sell{\'e}s},
  {Sahlmann}, {Salguero}, {Samaras}, {Sanchez Gimenez}, {Sanna},
  {Santove{\~n}a}, {Sarasso}, {Schultheis}, {Sciacca}, {Segol}, {Segovia},
  {S{\'e}gransan}, {Semeux}, {Shahaf}, {Siddiqui}, {Siebert}, {Siltala},
  {Silvelo}, {Slezak}, {Slezak}, {Smart}, {Snaith}, {Solano}, {Solitro},
  {Souami}, {Souchay}, {Spagna}, {Spina}, {Spoto}, {Steele},
  {Steidelm{\"u}ller}, {Stephenson}, {S{\"u}veges}, {Surdej}, {Szabados},
  {Szegedi-Elek}, {Taris}, {Taylo}, {Teixeira}, {Tolomei}, {Tonello}, {Torra},
  {Torra}, {Torralba Elipe}, {Trabucchi}, {Tsounis}, {Turon}, {Ulla}, {Unger},
  {Vaillant}, {van Dillen}, {van Reeven}, {Vanel}, {Vecchiato}, {Viala},
  {Vicente}, {Voutsinas}, {Weiler}, {Wevers}, {Wyrzykowski}, {Yoldas}, {Yvard},
  {Zhao}, {Zorec}, {Zucker}, \& {Zwitter}}]{2022arXiv220800211G}
{Gaia Collaboration}, {Vallenari}, A., {Brown}, A.~G.~A., {et~al.} 2022, arXiv
  e-prints, arXiv:2208.00211

\bibitem[{{Galli} {et~al.}(2019){Galli}, {Loinard}, {Bouy}, {Sarro},
  {Ortiz-Le{\'o}n}, {Dzib}, {Olivares}, {Heyer}, {Hernandez},
  {Rom{\'a}n-Z{\'u}{\~n}iga}, {Kounkel}, \& {Covey}}]{2019A&A...630A.137G}
{Galli}, P.~A.~B., {Loinard}, L., {Bouy}, H., {et~al.} 2019, \aap, 630, A137

\bibitem[{{Garufi} {et~al.}(2021){Garufi}, {Podio}, {Codella}, {Fedele},
  {Bianchi}, {Favre}, {Bacciotti}, {Ceccarelli}, {Mercimek}, {Rygl}, {Teague},
  \& {Testi}}]{2021A&A...645A.145G}
{Garufi}, A., {Podio}, L., {Codella}, C., {et~al.} 2021, \aap, 645, A145

\bibitem[{{Goldsmith} {et~al.}(2008){Goldsmith}, {Heyer}, {Narayanan}, {Snell},
  {Li}, \& {Brunt}}]{2008ApJ...680..428G}
{Goldsmith}, P.~F., {Heyer}, M., {Narayanan}, G., {et~al.} 2008, \apj, 680, 428

\bibitem[{{Gomez de Castro} \& {Pudritz}(1992)}]{1992ApJ...395..501G}
{Gomez de Castro}, A. \& {Pudritz}, R.~E. 1992, \apj, 395, 501

\bibitem[{{G{\'o}mez de Castro} {et~al.}(2015){G{\'o}mez de Castro},
  {Lopez-Santiago}, {L{\'o}pez-Mart{\'\i}nez}, {S{\'a}nchez}, {Sestito}, {de
  Castro}, {Cornide}, \& {Ya{\~n}ez Gestoso}}]{2015ApJS..216...26G}
{G{\'o}mez de Castro}, A.~I., {Lopez-Santiago}, J., {L{\'o}pez-Mart{\'\i}nez},
  F., {et~al.} 2015, \apjs, 216, 26

\bibitem[{{Herbig} \& {Kameswara Rao}(1972)}]{1972ApJ...174..401H}
{Herbig}, G.~H. \& {Kameswara Rao}, N. 1972, \apj, 174, 401

\bibitem[{{Janson} {et~al.}(2013){Janson}, {Brandt}, {Moro-Mart{\'\i}n},
  {Usuda}, {Thalmann}, {Carson}, {Goto}, {Currie}, {McElwain}, {Itoh},
  {Fukagawa}, {Crepp}, {Kuzuhara}, {Hashimoto}, {Kudo}, {Kusakabe}, {Abe},
  {Brandner}, {Egner}, {Feldt}, {Grady}, {Guyon}, {Hayano}, {Hayashi},
  {Hayashi}, {Henning}, {Hodapp}, {Ishii}, {Iye}, {Kandori}, {Knapp}, {Kwon},
  {Matsuo}, {Miyama}, {Morino}, {Nishimura}, {Pyo}, {Serabyn}, {Suenaga},
  {Suto}, {Suzuki}, {Takahashi}, {Takami}, {Takato}, {Terada}, {Tomono},
  {Turner}, {Watanabe}, {Wisniewski}, {Yamada}, {Takami}, \&
  {Tamura}}]{2013ApJ...773...73J}
{Janson}, M., {Brandt}, T.~D., {Moro-Mart{\'\i}n}, A., {et~al.} 2013, \apj,
  773, 73

\bibitem[{{Joncour} {et~al.}(2017){Joncour}, {Duch{\^e}ne}, \&
  {Moraux}}]{2017A&A...599A..14J}
{Joncour}, I., {Duch{\^e}ne}, G., \& {Moraux}, E. 2017, \aap, 599, A14

\bibitem[{{Kohoutek} \& {Wehmeyer}(1999)}]{1999A&AS..134..255K}
{Kohoutek}, L. \& {Wehmeyer}, R. 1999, \aaps, 134, 255

\bibitem[{Lloyd(2006)}]{10.1109/TIT.1982.1056489}
Lloyd, S. 2006, IEEE Trans. Inf. Theor., 28, 129–137

\bibitem[{{Lombardi} {et~al.}(2010){Lombardi}, {Lada}, \&
  {Alves}}]{2010A&A...512A..67L}
{Lombardi}, M., {Lada}, C.~J., \& {Alves}, J. 2010, \aap, 512, A67

\bibitem[{{Long} {et~al.}(2019){Long}, {Herczeg}, {Harsono}, {Pinilla},
  {Tazzari}, {Manara}, {Pascucci}, {Cabrit}, {Nisini}, {Johnstone}, {Edwards},
  {Salyk}, {Menard}, {Lodato}, {Boehler}, {Mace}, {Liu}, {Mulders}, {Hendler},
  {Ragusa}, {Fischer}, {Banzatti}, {Rigliaco}, {van de Plas}, {Dipierro},
  {Gully-Santiago}, \& {Lopez-Valdivia}}]{2019ApJ...882...49L}
{Long}, F., {Herczeg}, G.~J., {Harsono}, D., {et~al.} 2019, \apj, 882, 49

\bibitem[{{Luhman}(2018)}]{2018AJ....156..271L}
{Luhman}, K.~L. 2018, \aj, 156, 271

\bibitem[{{Luhman}(2022)}]{2022arXiv221109785L}
{Luhman}, K.~L. 2022, arXiv e-prints, arXiv:2211.09785

\bibitem[{{Luhman}(2023)}]{2023AJ....165...37L}
{Luhman}, K.~L. 2023, \aj, 165, 37

\bibitem[{{Luhman} {et~al.}(2010){Luhman}, {Allen}, {Espaillat}, {Hartmann}, \&
  {Calvet}}]{2010ApJS..186..111L}
{Luhman}, K.~L., {Allen}, P.~R., {Espaillat}, C., {Hartmann}, L., \& {Calvet},
  N. 2010, \apjs, 186, 111

\bibitem[{{Luhman} {et~al.}(2017){Luhman}, {Mamajek}, {Shukla}, \&
  {Loutrel}}]{2017AJ....153...46L}
{Luhman}, K.~L., {Mamajek}, E.~E., {Shukla}, S.~J., \& {Loutrel}, N.~P. 2017,
  \aj, 153, 46

\bibitem[{{Martin} {et~al.}(2005){Martin}, {Fanson}, {Schiminovich},
  {Morrissey}, {Friedman}, {Barlow}, {Conrow}, {Grange}, {Jelinsky},
  {Milliard}, {Siegmund}, {Bianchi}, {Byun}, {Donas}, {Forster}, {Heckman},
  {Lee}, {Madore}, {Malina}, {Neff}, {Rich}, {Small}, {Surber}, {Szalay},
  {Welsh}, \& {Wyder}}]{2005ApJ...619L...1M}
{Martin}, D.~C., {Fanson}, J., {Schiminovich}, D., {et~al.} 2005, \apjl, 619,
  L1

\bibitem[{{Mart{\'\i}n} {et~al.}(2010){Mart{\'\i}n}, {Phan-Bao}, {Bessell},
  {Delfosse}, {Forveille}, {Magazz{\`u}}, {Reyl{\'e}}, {Bouy}, \&
  {Tata}}]{2010A&A...517A..53M}
{Mart{\'\i}n}, E.~L., {Phan-Bao}, N., {Bessell}, M., {et~al.} 2010, \aap, 517,
  A53

\bibitem[{{Meshkat} {et~al.}(2017){Meshkat}, {Mawet}, {Bryan}, {Hinkley},
  {Bowler}, {Stapelfeldt}, {Batygin}, {Padgett}, {Morales}, {Serabyn},
  {Christiaens}, {Brandt}, \& {Wahhaj}}]{2017AJ....154..245M}
{Meshkat}, T., {Mawet}, D., {Bryan}, M.~L., {et~al.} 2017, \aj, 154, 245

\bibitem[{{Narayanan} {et~al.}(2008){Narayanan}, {Heyer}, {Brunt}, {Goldsmith},
  {Snell}, \& {Li}}]{2008ApJS..177..341N}
{Narayanan}, G., {Heyer}, M.~H., {Brunt}, C., {et~al.} 2008, \apjs, 177, 341

\bibitem[{{Neuhaeuser} {et~al.}(1995){Neuhaeuser}, {Sterzik}, {Schmitt},
  {Wichmann}, \& {Krautter}}]{1995A&A...297..391N}
{Neuhaeuser}, R., {Sterzik}, M.~F., {Schmitt}, J.~H.~M.~M., {Wichmann}, R., \&
  {Krautter}, J. 1995, \aap, 297, 391

\bibitem[{Pedregosa {et~al.}(2011)Pedregosa, Varoquaux, Gramfort, Michel,
  Thirion, Grisel, Blondel, Prettenhofer, Weiss, Dubourg, Vanderplas, Passos,
  Cournapeau, Brucher, Perrot, \& Duchesnay}]{scikit-learn}
Pedregosa, F., Varoquaux, G., Gramfort, A., {et~al.} 2011, Journal of Machine
  Learning Research, 12, 2825

\bibitem[{{Planck Collaboration} {et~al.}(2016){Planck Collaboration}, {Adam},
  {Ade}, {Alves}, {Ashdown}, {Aumont}, {Baccigalupi}, {Banday}, {Barreiro},
  {Bartolo}, {Battaner}, {Benabed}, {Benoit-L{\'e}vy}, {Bernard}, {Bersanelli},
  {Bielewicz}, {Bonavera}, {Bond}, {Borrill}, {Bouchet}, {Boulanger}, {Bucher},
  {Burigana}, {Butler}, {Calabrese}, {Cardoso}, {Catalano}, {Chiang},
  {Christensen}, {Colombo}, {Combet}, {Couchot}, {Crill}, {Curto}, {Cuttaia},
  {Danese}, {Davis}, {de Bernardis}, {de Rosa}, {de Zotti}, {Delabrouille},
  {Dickinson}, {Diego}, {Dolag}, {Dor{\'e}}, {Ducout}, {Dupac}, {Elsner},
  {En{\ss}lin}, {Eriksen}, {Ferri{\`e}re}, {Finelli}, {Forni}, {Frailis},
  {Fraisse}, {Franceschi}, {Galeotta}, {Ganga}, {Ghosh}, {Giard}, {Gjerl{\o}w},
  {Gonz{\'a}lez-Nuevo}, {G{\'o}rski}, {Gregorio}, {Gruppuso}, {Gudmundsson},
  {Hansen}, {Harrison}, {Hern{\'a}ndez-Monteagudo}, {Herranz}, {Hildebrandt},
  {Hobson}, {Hornstrup}, {Hurier}, {Jaffe}, {Jaffe}, {Jones}, {Juvela},
  {Keih{\"a}nen}, {Keskitalo}, {Kisner}, {Knoche}, {Kunz}, {Kurki-Suonio},
  {Lamarre}, {Lasenby}, {Lattanzi}, {Lawrence}, {Leahy}, {Leonardi}, {Levrier},
  {Liguori}, {Lilje}, {Linden-V{\o}rnle}, {L{\'o}pez-Caniego}, {Lubin},
  {Mac{\'\i}as-P{\'e}rez}, {Maggio}, {Maino}, {Mandolesi}, {Mangilli}, {Maris},
  {Martin}, {Mart{\'\i}nez-Gonz{\'a}lez}, {Masi}, {Matarrese}, {Melchiorri},
  {Mennella}, {Migliaccio}, {Miville-Desch{\^e}nes}, {Moneti}, {Montier},
  {Morgante}, {Munshi}, {Murphy}, {Naselsky}, {Nati}, {Natoli},
  {N{\o}rgaard-Nielsen}, {Oppermann}, {Orlando}, {Pagano}, {Pajot}, {Paladini},
  {Paoletti}, {Pasian}, {Perotto}, {Pettorino}, {Piacentini}, {Piat},
  {Pierpaoli}, {Plaszczynski}, {Pointecouteau}, {Polenta}, {Ponthieu}, {Pratt},
  {Prunet}, {Puget}, {Rachen}, {Reinecke}, {Remazeilles}, {Renault}, {Renzi},
  {Ristorcelli}, {Rocha}, {Rossetti}, {Roudier}, {Rubi{\~n}o-Mart{\'\i}n},
  {Rusholme}, {Sandri}, {Santos}, {Savelainen}, {Scott}, {Spencer},
  {Stolyarov}, {Stompor}, {Strong}, {Sudiwala}, {Sunyaev}, {Suur-Uski},
  {Sygnet}, {Tauber}, {Terenzi}, {Toffolatti}, {Tomasi}, {Tristram}, {Tucci},
  {Valenziano}, {Valiviita}, {Van Tent}, {Vielva}, {Villa}, {Wade}, {Wandelt},
  {Wehus}, {Yvon}, {Zacchei}, \& {Zonca}}]{2016A&A...596A.103P}
{Planck Collaboration}, {Adam}, R., {Ade}, P.~A.~R., {et~al.} 2016, \aap, 596,
  A103

\bibitem[{{Roccatagliata} {et~al.}(2020){Roccatagliata}, {Franciosini},
  {Sacco}, {Randich}, \& {Sicilia-Aguilar}}]{2020A&A...638A..85R}
{Roccatagliata}, V., {Franciosini}, E., {Sacco}, G.~G., {Randich}, S., \&
  {Sicilia-Aguilar}, A. 2020, \aap, 638, A85

\bibitem[{{Stassun} \& {Torres}(2021)}]{2021ApJ...907L..33S}
{Stassun}, K.~G. \& {Torres}, G. 2021, \apjl, 907, L33

\bibitem[{{Ungerechts} \& {Thaddeus}(1987)}]{1987ApJS...63..645U}
{Ungerechts}, H. \& {Thaddeus}, P. 1987, \apjs, 63, 645

\bibitem[{{Wichmann} {et~al.}(2000){Wichmann}, {Torres}, {Melo}, {Frink},
  {Allain}, {Bouvier}, {Krautter}, {Covino}, \&
  {Neuh{\"a}user}}]{2000A&A...359..181W}
{Wichmann}, R., {Torres}, G., {Melo}, C.~H.~F., {et~al.} 2000, \aap, 359, 181

\bibitem[{{Xing}(2010)}]{2010ApJ...723.1542X}
{Xing}, L.~F. 2010, \apj, 723, 1542

\end{thebibliography}
%
\begin{appendix}

\section{Impact of uncertainties on our clusterings analysis}\label{appendix_RFM}
         In order to account for the data uncertainties, we used a Monte Carlo sampling technique to generate 10$^{4}$ synthetic datasets.
         For each one, we applied the k-means++ algorithm, keeping track of the number of groups generated and the positions of the centroids
         (the nearest neighbour was used as proxy to obtain physical coordinates).
         Therefore, for each source in our list of bona-fide TTSs, we generated 10$^{4}$ realizations of its astrometric parameters.
         {\it Gaia}~DR3 provides the correlations between each pair of astrometric parameters. Given two parameters, $x$ and $y$, with
         standard deviations $\sigma_{x}$ and $\sigma_{y}$, their respective covariance and correlation coefficients, $\sigma_{xy}$ and
         $\rho_{xy}$, can be written as $\sigma_{xy}=\rho_{xy}\,\sigma_{x}\,\sigma_{y}$. If \textbf{\textsf{C}} is the covariance matrix
         at a given epoch associated with the astrometric solution that is symmetric and positive-semidefinite, then
         \textbf{\textsf{C}} = \textbf{\textsf{A}} \textbf{\textsf{A}}$^{\textbf{\textsf{T}}}$ where \textbf{\textsf{A}} is a lower
         triangular matrix with real and positive diagonal elements, \textbf{\textsf{A}}$^{\textbf{\textsf{T}}}$ is the transpose of
         \textbf{\textsf{A}}. In the particular case studied here, these matrices are $5\times5$. If the elements of \textbf{\textsf{C}} are
         written as $c_{\rm ij}=\rho_{\rm ij}\,\sigma_{\rm i}\,\sigma_{\rm j}$ and those of \textbf{\textsf{A}} as $a_{\rm ij}$, where
         those are the entries in the $i$-th row and $j$-th column, and if $\vec{r}$ is a vector made of univariate Gaussian random numbers
         (components $r_{i}$ with $i=1,5$), the required multivariate Gaussian random samples are given by the expressions:
         \begin{equation}
            \begin{aligned}
               \alpha_{\rm c}        & = \alpha + a_{11}\,r_{1} \\
               \delta_{\rm c}        & = \delta + a_{22}\,r_{2} + a_{21}\,r_{1} \\
               \pi_{\rm c}           & = \pi + a_{33}\,r_{3} + a_{32}\,r_{2} + a_{31}\,r_{1} \\
               \mu_{\alpha\ {\rm c}} & = \mu_{\alpha} + a_{44}\,r_{4} + a_{43}\,r_{3} + a_{42}\,r_{2} + a_{41}\,r_{1} \\
               \mu_{\delta\ {\rm c}} & = \mu_{\delta} + a_{55}\,r_{5} + a_{54}\,r_{4} + a_{53}\,r_{3} + a_{52}\,r_{2} + a_{51}\,r_{1}  \,,
               \label{cova}
            \end{aligned}
         \end{equation}
         where $\alpha$, $\delta$, $\pi$, $\mu_{\alpha}$, and $\mu_{\delta}$ are the values of right ascension, declination, absolute stellar
         parallax and proper motions in right ascension and declination directions provided by {\it Gaia}~DR3 and the $a_{ij}$ coefficients are
         given by:
         \begin{equation}
            \begin{aligned}
               a_{11} & = \sigma_{\alpha} \\
               a_{21} & = \rho_{\alpha\ \delta}\,\sigma_{\delta} \\
               a_{31} & = \rho_{\alpha\ \pi}\,\sigma_{\pi} \\
               a_{41} & = \rho_{\alpha\ \mu_{\alpha}}\,\sigma_{\mu_{\alpha}} \\
               a_{51} & = \rho_{\alpha\ \mu_{\delta}}\,\sigma_{\mu_{\delta}} \\
               a_{22} & = \sqrt{\sigma_{\delta}^{2} - a_{21}^{2}} \\
               a_{32} & = (\rho_{\delta\ \pi}\,\sigma_{\delta}\,\sigma_{\pi} - a_{21}\,a_{31}) / a_{22} \\
               a_{42} & = (\rho_{\delta\ \mu_{\alpha}}\,\sigma_{\delta}\,\sigma_{\mu_{\alpha}} - a_{21}\,a_{41}) / a_{22} \\
               a_{52} & = (\rho_{\delta\ \mu_{\delta}}\,\sigma_{\delta}\,\sigma_{\mu_{\delta}} - a_{21}\,a_{51}) / a_{22} \\
               a_{33} & = \sqrt{\sigma_{\pi}^{2} - a_{31}^{2} - a_{32}^{2}} \\
               a_{43} & = (\rho_{\pi\ \mu_{\alpha}}\,\sigma_{\pi}\,\sigma_{\mu_{\alpha}} - a_{31}\,a_{41} - a_{32}\,a_{42}) / a_{33} \\
               a_{53} & = (\rho_{\pi\ \mu_{\delta}}\,\sigma_{\pi}\,\sigma_{\mu_{\delta}} - a_{31}\,a_{51} - a_{32}\,a_{52}) / a_{33} \\
               a_{44} & = \sqrt{\sigma_{\mu_{\alpha}}^{2} - a_{41}^{2} - a_{42}^{2} - a_{43}^{2}} \\
               a_{54} & = (\rho_{\mu_{\alpha}\ \mu_{\delta}}\,\sigma_{\mu_{\alpha}}\,\sigma_{\mu_{\delta}} - a_{41}\,a_{51} - a_{42}\,a_{52} - a_{43}\,a_{53}) / a_{44} \\
               a_{55} & = \sqrt{\sigma_{\mu_{\delta}}^{2} - a_{51}^{2} - a_{52}^{2} - a_{53}^{2} - a_{54}^{2}} \,,
            \end{aligned}
         \end{equation}
         where $\sigma_{\alpha}$, $\sigma_{\delta}$, $\sigma_{\pi}$, $\sigma_{\mu_{\alpha}}$, and $\sigma_{\mu_{\delta}}$ are the standard errors in
         right ascension, declination, parallax and proper motions from {\it Gaia}~DR3, and $\rho_{\alpha\ \delta}$, $\rho_{\alpha\ \pi}$,
         $\rho_{\alpha\ \mu_{\alpha}}$, $\rho_{\alpha\ \mu_{\delta}}$, $\rho_{\delta\ \pi}$, $\rho_{\delta\ \mu_{\alpha}}$, $\rho_{\delta\ \mu_{\delta}}$,
         $\rho_{\pi\ \mu_{\alpha}}$, $\rho_{\pi\ \mu_{\delta}}$, and $\rho_{\mu_{\alpha}\ \mu_{\delta}}$ their
         respective correlation coefficients, also from {\it Gaia}~DR3. For each $\pi_{\rm c}$, we computed the value of the distance by
         applying the usual relationship, $d_{\rm c}=1/\pi_{\rm c}$.
  
\begin{figure}
\includegraphics[width=\linewidth]{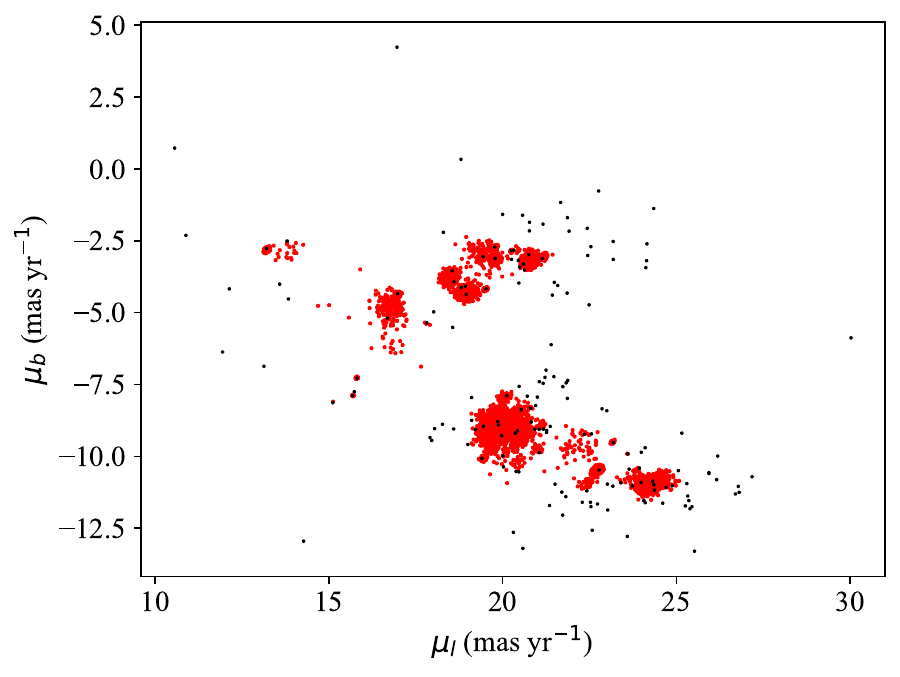}
\caption{Effect of the uncertainties on the clustering analysis. Results in physical parameters.
In black we have the sources and in red the 10$^{4}$ centroids.}
\label{MCCMPlot}
\end{figure}

  For each set of synthetic sources statistically compatible with the qualification sample, we
compute the proper motions along the Galactic coordinates as pointed out above and use their
values together with the parallaxes to apply the $k$-means++ algorithm. The resulting centroids
are shown in Fig.~\ref{MCCMPlot} in red.

\section{Fitting the logistic regression model with errors}\label{appendix_LBA}

\begin{figure*}
\includegraphics[width = \linewidth]{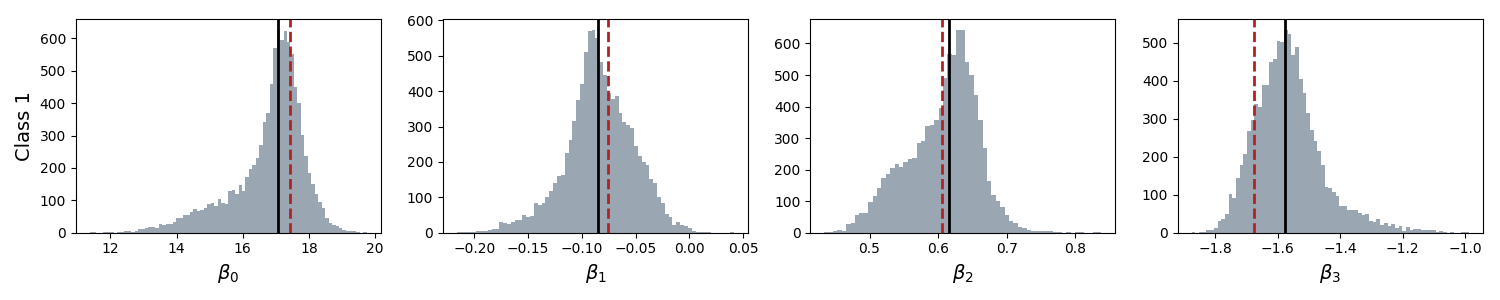}
\caption{Final distributions for the logistic regression coefficients 
after $10^{4}$ Monte Carlo realizations. The solid black line corresponds
to the median value adopted as the best-fit, while the red dashed line corresponds
to the value obtained from the fit of the original variables without taking errors into account.}
\label{fig:MC_LR_params}
\end{figure*}

From the mathematical point of view, the logistic regression model is trained
on a set of qualification sources with well-defined groups, where all the values
are considered to be free of errors. However, the proper motions and parallax of 
our training sample of bona fide TTS have inherent errors and depend one to each
other -- the parameters are correlated. In order to take this constraint into
account, we produce a set of $10^{4}$ mock realizations of the sample
using the covariance matrix for generating random samples of ($\mu_{\alpha ^{*}}, \mu_{\delta}, \pi$)
that we then convert into proper motions in Galactic coordinates as explained in the
\textit{Gaia} documentation\footnote{\url{https://gea.esac.esa.int/archive/documentation/GEDR3/Data_processing/chap_cu3ast/sec_cu3ast_intro/ssec_cu3ast_intro_tansforms.html}}. As a result, we obtain a
distribution for each value of the logistic regression model ($\beta_{m}$, $m = 0, 1, 2, 3$)
as shown in 
Fig. \ref{fig:MC_LR_params}. In these plots we have included the median
value for each coefficient in black, which is taken as the best-fit value, as well
as the results if we fit the original sample without taking the errors into account.

For the classification of the TTS candidates, we perform an analogous Monte
Carlo analysis and generate again 10$^{4}$ samples taking into account the
errors, and for each realization we assign a membership probability. The values
given in the main text are the median values of the probabilities of the
$10^{4}$ random samples.

\end{appendix}

\end{document}